\documentclass[10pt]{article}

\def\fssf{\footnotesize\sf}

\usepackage{epsfig}
\usepackage{amssymb}
\usepackage{amsmath}
\usepackage{verbatim}
\usepackage{url}

\renewcommand{\baselinestretch}{2.0}

\begin{document}

\pagestyle{plain}
\pagenumbering{arabic}
\setcounter{page}{1}

\vspace*{-2.5cm} \hspace*{10cm} Last Modified: 22 Sept. 2009

\begin{center}

{\Large \bf \bf Structural Consistency: Enabling XML Keyword Search\\ to Eliminate Spurious Results Consistently}

Ki-Hoon Lee$^{\dagger}$, Kyu-Young Whang$^{\dagger}$, Wook-Shin Han$^{\dagger\dagger}$, and Min-Soo Kim$^{\dagger}$\\
\vspace*{-0.20cm}
$^{\dagger}$Department of Computer Science\\
\vspace*{-0.20cm}
Korea Advanced Institute of Science and Technology\,(KAIST) \\
\vspace*{-0.20cm}
$^{\dagger\dagger}$Department of Computer Engineering\\
\vspace*{-0.20cm}
Kyungpook National University \\
\vspace*{-0.20cm}
e-mail:\,$^{\dagger}$\{khlee, kywhang, mskim\}@mozart.kaist.ac.kr, $^{\dagger\dagger}$wshan@knu.ac.kr\\
\end{center}


\renewcommand{\baselinestretch}{1.6}
\begin{abstract}
{\small
XML keyword search is a user-friendly way to query XML data using only keywords. In XML keyword search, to achieve high precision without sacrificing recall, it is important to remove {\it spurious} results not intended by the user. Efforts to eliminate spurious results have enjoyed some success by using the concepts of LCA or its variants, SLCA and MLCA. However, existing methods still could find many spurious results. The fundamental cause for the occurrence of spurious results is that the existing methods try to eliminate spurious results locally without global examination of all the query results and, accordingly, some spurious results are not consistently eliminated.
In this paper, we propose a novel keyword search method that removes spurious results consistently by exploiting the new concept of structural consistency. We define {\it structural consistency} as a property that is preserved if there is no query result having an ancestor-descendant relationship {\it at the schema level} with any other query results. A naive solution to obtain structural consistency would be to compute all the LCAs (or variants) and then to remove spurious results according to structural consistency. Obviously, this approach would always be slower than existing LCA-based ones. To speed up structural consistency checking, we must be able to examine the query results at the schema level without generating all the LCAs. However, this is a challenging problem since the schema-level query results do not homomorphically map to the instance-level query results, causing serious false dismissal. We present a comprehensive and practical solution to this problem and formally prove that this solution preserves structural consistency at the schema level without incurring false dismissal. We also propose a relevance-feedback based solution for the problem where our method has low recall, which occurs when it is not the user's intention to find more specific results. This solution has been prototyped in a full-fledged object-relational DBMS. Experimental results using real and synthetic data sets show that, compared with the state-of-the-art methods, our solution significantly 1) improves precision while providing comparable recall for most queries and 2) enhances the query performance by removing spurious results early. 
}
\end{abstract}
\renewcommand{\baselinestretch}{2.0}


\newtheorem{definition}{\bf Definition}
\newtheorem{strategy}{\bf Strategy}
\newtheorem{example}{\bf Example}
\newtheorem{property}{\bf Property}
\newtheorem{corollary}{\bf Corollary}
\newtheorem{theorem}{\bf Theorem}
\newtheorem{lemma}{\bf Lemma}

\newenvironment{newidth}[2]{
 \begin{list}{}{
  \setlength{\topsep}{0pt}
  \setlength{\leftmargin}{#1}
  \setlength{\rightmargin}{#2}
  \setlength{\listparindent}{\parindent}
  \setlength{\itemindent}{\parindent}
  \setlength{\parsep}{\parskip}
 }
\item[]}{\end{list}}

\section{Introduction}\label{sec:introduction}
As XML becomes the standard for data representation and exchange on
the Internet, querying XML data has become an important
issue\,\cite{Li08}. Research work in this area can be classified
into two categories: the structured query approach and the keyword
query approach\,\cite{Li08}. Both approaches have tradeoffs. The
 structured query approach specifies the precise
structure of the desired results using a structured query language
such as XPath and XQuery. However, it is hard to formulate queries
without prior knowledge about structured query languages or without
knowing the schema of the XML data. The keyword query, on the other
hand, can overcome this problem by requiring only keywords rather
than specific structure information. This approach, however, might
not deliver precise results since it does not contain precise
structures.

In the structured query, the user's query intention can be expressed
as either a single structured query or multiple structured queries,
depending on the heterogeneity of the underlying XML data. If there
is only one structure matching the user's intention at the schema
level, that intention can be expressed in a single structured query.
However, if there are multiple structures matching the user's
intention, multiple structured queries for those structures must be
composed.

\begin{figure*}[hbt]
\begin{center}
\begin{minipage}{10.8cm}
\begin{center}
\centerline{\includegraphics[width=13cm]
{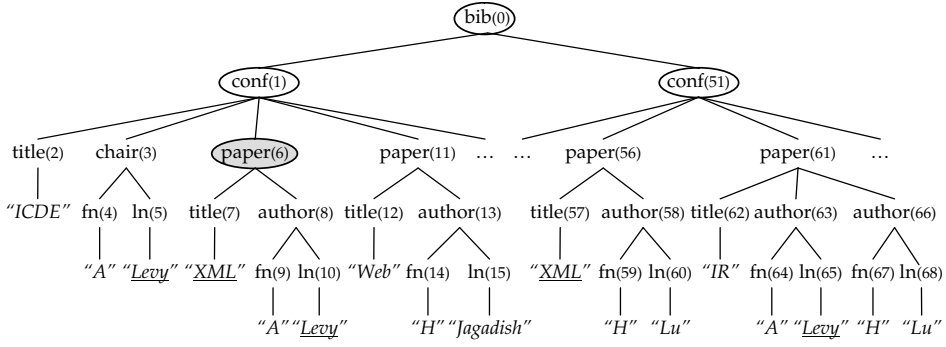}}
\ifx\vldbjformat\undefined \vspace{-0.8cm} \else \vspace{-0.5cm} \fi
{\footnotesize (a) XML data on conference publications.}
\end{center}
\end{minipage}
\begin{minipage}{7.2cm}
\begin{center}
\vspace{0.5cm}
\centerline{\includegraphics[width=9.03cm]
{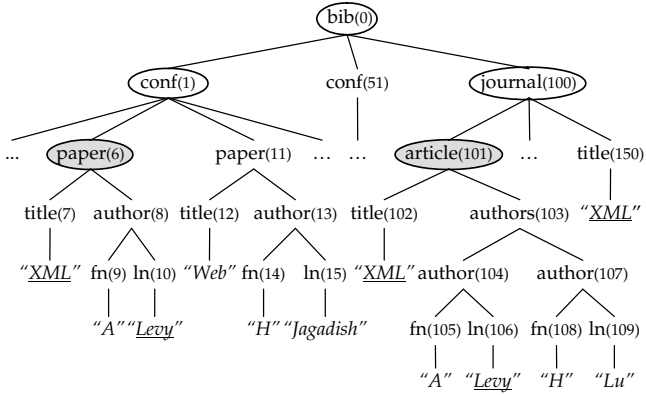}}
\vspace{0.1cm}
{\footnotesize (b) XML data on conference and journal publications.}
\end{center}
\end{minipage}
\end{center}
\ifx\vldbjformat\undefined \vspace{-0.2cm} \fi
\caption{Querying XML data.}
\label{fig:motivating_example}
\end{figure*}

\begin{example}\label{eg:motivating_example1}
{\rm The XML data in Fig.~\ref{fig:motivating_example}(a)
represent bibliographic data on conference publications. Suppose
that a user intends to find the publications of {\fssf``Levy"}
on {\fssf``XML"}. This query can be stated as a single
structured query, $Q_1$; in the keyword query, it is represented
as {\fssf``XML Levy"}. The query result is
\{{\fssf paper(6)}\}. Here, we denote the subtree rooted at
node $p$ as $p$ in the same way as is done by Xu and Papakonstantinou\,\cite{Xu05}.}

\vspace{0.1cm}

$Q_1$: {\fssf
/bib/conf/paper[``XML"][``Levy"]}\footnote{For ease of
exposition, we denote the predicate that checks whether a keyword {\sf w}
is contained in an element {\sf e} as {\sf e\,[``w"]} instead of
{\sf e\,[contains(., ``w")]} that uses the {\sf contains} function in the
XPath standard.}\hfill $\Box$
\end{example}

\begin{example}\label{eg:motivating_example2}
{\rm The XML data in Fig.~\ref{fig:motivating_example}(b)
represent bibliographic data on conference and journal
publications. Here, the subtree rooted at {\fssf conf(1)} is
the same as in Fig.~\ref{fig:motivating_example}(a). Since there
are two structures matching the user's intention, one for conference
papers and the other for journal articles, a union of multiple
structured queries, $Q_2$, must be used to find the desired results
despite the same query intention as in
Example~\ref{eg:motivating_example1}. Note that we still use the
same keyword query as in Example~\ref{eg:motivating_example1}. The
query results are \{{\fssf paper(6)}, {\fssf article(101)}\}.

\vspace{0.1cm}

$Q_2$: {\fssf /bib/conf/paper[``XML"][``Levy"] union }

\ifx\vldbjformat\undefined \vspace{-0.5cm} \fi

\hspace*{0.55cm}
{\fssf /bib/journal/article[``XML"][``Levy"]}}\hfill$\Box$
\end{example}

In the keyword search, a user wants to have high recall and high
precision\,\cite{Baez99}. A naive way to achieve high recall (100\%) in
XML keyword search would be to return the root of an XML document.
However, with this approach, the user would suffer from very low
precision due to a large amount of spurious results not intended
by the user.

Efforts to eliminate spurious results\,\cite{Cohe03,Guo03,Li08,Xu05} have enjoyed some success
by using the concepts of LCA or its variants, SLCA\,\cite{Xu05} and
MLCA\,\cite{Li08}. For a keyword query $Q$\,=\,\{$w_1$,\,$w_2$,\,..., $w_m$\}, an LCA is the common ancestor node of nodes $n_1$, $n_2$, ..., $n_m$ where $n_i$ is a node directly containing $w_i$ ($1$$\leq$$i$$\leq$$m$). It is located farthest from
the root node. The SLCA method, a refinement of the LCA method, finds
LCAs that do not contain other LCAs. For example, if we use the LCA
method to find the results in Fig.~\ref{fig:motivating_example}(a),
\{{\fssf bib(0)}, {\fssf conf(1)}, {\fssf paper(6)}, {\fssf conf(51)}\} are retrieved. With the SLCA me-thod, \{{\fssf paper(6)}, {\fssf conf(51)}\} are retrieved. As shown here, existing methods for XML keyword search still could find many {\it spurious} results\,(e.g., \{{\fssf bib(0)},
{\fssf conf(1)}, {\fssf conf(51)}\}), i.e., those that are not intended by the user. Here, following the common practice\,\cite{Cohe03,Li07,Li08}, we define {\it correct} results of a keyword query as those returned by structured queries (such as $Q_1$) corresponding to the keyword query, which are formulated according to the schema of the underlying XML data. In the real data set (DBLP), spurious results such as {\fssf conf(51)} can include huge subtrees having
thousands of nodes. This serious problem of low precision in the-state-of-art methods not only overburdens the user with filtering
numerous spurious results, but also degrades the performance of the
system due to unnecessary computation. For instance, if we issue a
keyword query {\fssf``XML Levy"} over the DBLP data set, we
obtain 388,066 nodes using the SLCA method, among which only 69 nodes
(precision = $\frac{69}{388,066}$ $\approx$ 0.02\%) are correct
results.

\begin{figure}[h]
\begin{center}
\begin{minipage}{4cm}
\begin{center}
\centerline{\includegraphics[width=3cm]
{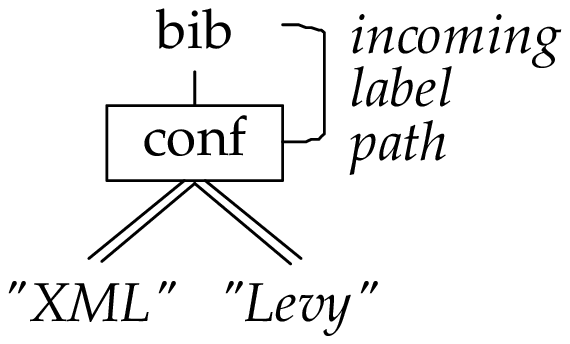}}
\ifx\vldbjformat\undefined \vspace{-0.1cm} \fi
{\footnotesize (a) $rs$({\sf conf(51)})}.
\end{center}
\end{minipage}
\begin{minipage}{4cm}
\begin{center}
\centerline{\includegraphics[width=3cm]
{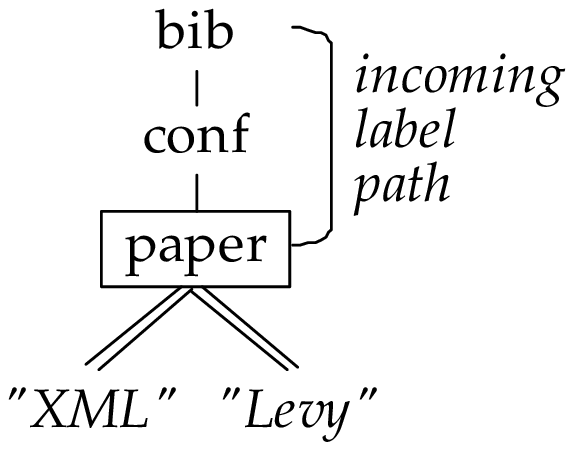}}
\ifx\vldbjformat\undefined \vspace{-0.1cm} \fi
{\footnotesize (b) $rs$({\sf paper(6)})}.
\end{center}
\end{minipage}
\end{center}
\vspace{-0.2cm}
\caption{The result structures of query results.}
\label{fig:smallest_query_structure}
\ifx\vldbjformat\undefined \vspace{-0.2cm} \fi
\end{figure}

The fundamental cause for the occurrence of spurious results is that the existing methods try to eliminate spurious results locally without global examination of all the query results. For instance, in Example~\ref{eg:motivating_example1}, the LCA method finds a correct result \{{\fssf paper(6)}\}, but also finds spurious results
\{{\fssf bib(0)}, {\fssf conf(1)}, {\fssf
conf(51)}\}. With the SLCA method, we can eliminate two spurious results \{{\fssf bib(0)}, {\fssf conf(1)}\} since
they contain other LCAs. However, {\fssf conf(51)}
still remains since it is not an ancestor of {\fssf paper(6)}. This is {\it inconsistent} since both {\fssf conf(1)} and {\fssf conf(51)} are spurious results having an identical result structure. Here, we define the {\it result structure}\footnote{Intuitively, the result structure is the schema of a query result (an instance).} of a query
result $qr$ as a (schema-level) twig pattern composed of the label path\,\cite{Gold97} from the root of the XML data to the root $qr_{root}$ of
$qr$\,(simply, the {\it incoming label path}) and the ancestor-descendant
edges from $qr_{root}$ to query keywords. In the result structure of a
query result $qr$, denoted by $rs(qr)$, the node corresponding to $qr_{root}$ is marked as the {\it query result node}\,\cite{Park05} and is distinguished from other nodes by placing it in a box.
Fig.~\ref{fig:smallest_query_structure} shows $rs$({\fssf
conf(51)}) and $rs$({\fssf paper(6)}).

\ifx\vldbjformat\undefined
\newpage
\fi

We observe that, if two query results have an ancestor-descendant
relationship {\it at the schema level}, the ancestor is spurious. We
call this phenomenon {\it structural anomaly}. Here, a query result
$qr_1$ is an ancestor of a query result $qr_2$ at the schema level if
and only if the incoming label path of $rs(qr_1)$ is a proper prefix
of that of $rs(qr_2)$. By examining the query results at the schema level, we can remove spurious results having the same result structure consistently.
For example, in Fig.~\ref{fig:motivating_example}(a), the query results of the SLCA method are \{{\fssf paper(6)}, {\fssf conf(51)}\}, and the incoming label path of $rs$({\fssf conf(51)}) is a proper prefix of that of $rs$({\fssf paper(6)}) as in Fig.~\ref{fig:smallest_query_structure}. Hence, {\fssf conf(51)}, which has the same result structure as {\fssf conf(1)}, is spurious.

We argue that, to improve precision, there should be no structural anomaly in the query results. We call this property {\it structural consistency} (to be defined more formally in Section~\ref{sec:concept}). Otherwise, we are bound to retrieve inconsistent spurious results.

In this paper, we resolve structural anomalies by exploiting the
notion of the smallest result structure. The {\it smallest result
structure} is defined to be a result structure whose incoming label path is not a proper prefix of those of any other result structures.
We then remove the query result whose structure is not the same as a
smallest result structure, thereby obtaining structural consistency. For example, the smallest result structure of \{{\fssf paper(6)}, {\fssf conf(51)}\} is $rs$({\fssf paper(6)}) in Fig. \ref{fig:smallest_query_structure}(b) since the
incoming label path of $rs$({\fssf paper(6)}) is not a prefix of
that of $rs$({\fssf conf(51)}). Thus, {\fssf conf(51)} is removed.

A naive instance-level approach to obtain structural consistency would be to compute all the LCAs (or variants) and then to remove spurious results according to structural consistency. Obviously, this approach would always be slower than existing LCA-based ones. To speed up structural consistency checking, we must examine the query results at the schema level without generating all the LCAs.

The challenging issue here is ``How do we formally guarantee that the schema-level approach produces the same query results as the instance-level approach does?'' That is, if we blindly find SLCAs at the schema
level and compute answers using the SLCAs, we may encounter a {\it false dismissal} problem (to be elaborated in more detail in Section~\ref{sec:relationship}). For example, an empty result can be obtained even though query results corresponding to smallest result structures exist as in Example~\ref{eg:false_dismissal}. We may also encounter {\it phantom schema-level SLCAs} (to be defined in Section~\ref{sec:relationship}), which incurs structural anomaly. These problems occur because the
schema-level SLCAs do not homomorphically map to the instance-level
SLCAs. As a solution to these problems, we introduce the concept of
{\it iterative $k$th-ancestor generalization}, which iteratively finds
the $k$th-ancestors of SLCAs at the schema level and removes phantom schema-level SLCAs. Through iterative $k$th-ancestor generalization, the schema-level definition of structural consistency becomes equivalent to the instance-level one, and we formally prove this equivalence in Theorem~\ref{theorem:schema_level} of Section~\ref{sec:schema_algorithm}.

\begin{figure}[h]
\vspace{-0.6cm}
\begin{center}
\begin{minipage}{4cm}
\begin{center}
\centerline{\includegraphics[width=1.7cm]
{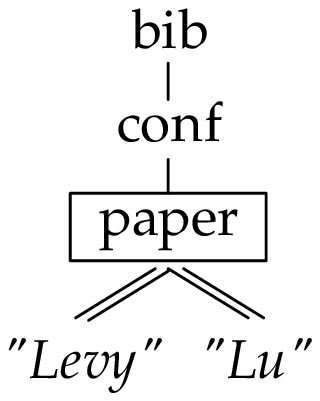}}
{\footnotesize (a) The smallest structure.}
\end{center}
\end{minipage}
\begin{minipage}{4cm}
\begin{center}
\centerline{\includegraphics[width=1.7cm]
{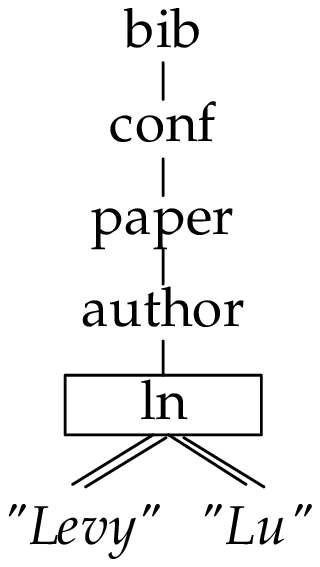}}
{\footnotesize (b) The XPath query obtained from the schema-level SLCA.}
\end{center}
\end{minipage}
\end{center}
\ifx\vldbjformat\undefined \vspace{-0.2cm} \fi
\caption{An example of false dismissal.}
\label{fig:Levy_Lu_Structure1}
\end{figure}

\begin{example}\label{eg:false_dismissal}
{\rm
Consider a keyword query $Q$ = \{{\fssf ``Levy"}, {\fssf ``Lu"}\} issued on the XML data in Fig.~\ref{fig:motivating_example}(a).
In the XML data in Fig.~\ref{fig:motivating_example}(a), we see that there is a query result, {\fssf paper(61)}, corresponding to the smallest result structure shown in Fig.~\ref{fig:Levy_Lu_Structure1}(a). However, there is no query result corresponding to the XPath query shown in Fig.~\ref{fig:Levy_Lu_Structure1}(b) that is obtained from the schema-level SLCA. (We will formally define the schema-level SLCA in Section~\ref{sec:schema_slca}.)}\hfill $\Box$
\end{example}

The contributions of this paper are as follows: 1) we formally propose new notions of structural consistency and structural anomaly; 2) we formally analyze the relationship between the set of schema-level SLCAs and the set of instance-level SLCAs, and then, propose an efficient algorithm that resolves structural anomaly at the schema level using the relationship analyzed. (we call this algorithm {\it schema-level structural anomaly resolution}.); 3) we formally prove in Theorem~\ref{theorem:schema_level} that this algorithm preserves structural consistency as is originally defined at the instance-level without incurring false dismissal; 4) we propose a relevance-feedback base solution for the problem where our method has low recall, which occurs when it is not the user's intention to find more specific results.; 5) we propose an efficient algorithm that simultaneously evaluates the multiple XPath queries generated by our method; 6) we have prototyped this algorithm in a full-fledged object-relational DBMS\,\cite{Whan05}; 7) we perform extensive experiments using real and synthetic data sets. The results show that we can significantly reduce spurious results compared with the existing methods by exploiting structural consistency. Furthermore, the experimental results show that our schema-level algorithm significantly improves the query performance over the existing ones.

The rest of this paper is organized as follows. Section~\ref{sec:background} describes the XML data model, schema of XML data, query models, and quality measure of XML keyword search. Section~\ref{sec:structural_consistency} proposes the concept of structural consistency and schema-level structural anomaly resolution. Section~\ref{sec:implementation} presents the implementation of schema-level structural anomaly resolution. Section~\ref{sec:related_work} reviews existing work, and Section~\ref{sec:experiments} presents the experimental results.
Finally, Section~\ref{sec:conclusions} presents our conclusions.

\section{Background}
\label{sec:background}
\subsection{XML Data Model}
We model XML data as a labeled tree\,\cite{Cohe03,Li08,Liu07,Xu05} where a node represents an element, attribute, or value, and
an edge represents the parent-child relationship between two nodes.
Every element or attribute node has a {\it label} and a unique {\it
id}, and each id is assigned a preorder number. A node that has a
label {\it l} and an id {\it i} is denoted as {\it l}({\it i}).
Definition~\ref{def:label_path} defines the label path of a node, and Definition~\ref{def:node_path} the node path.

\begin{definition}{\rm\cite{Gold97}\label{def:label_path}
The {\it label path} of a node $o$ is defined as a sequence of node
labels $l_1, l_2, ..., l_m$ from the root to the node $o$, and is
denoted as $l_1.l_2.\cdots.l_m$.}\hfill$\Box$
\end{definition}
\ifx\vldbjformat\undefined \vspace{-0.5cm} \fi
\begin{definition}{\rm\cite{Park05}\label{def:node_path}
The {\it node path} of a node $o$ is defined as a sequence of node
identifiers $n_1, n_2, ..., n_m$ from the root to the node $o$, and is
denoted as $n_1.n_2.\cdots.n_m$. We denote the $i$th id of a node path $node\_path$ as $node\_path[i]$. We note that the ids $n_1, n_2, ..., n_m$ have an ascending order since each $n_i$ $(1$$\leq$$i$$\leq$$m)$ is assigned a preorder number.
}\hfill$\Box$
\end{definition}

\subsection{Schema of XML Data}
Although DTD or XML Schema are used as the schema of XML data, XML
data often do not have them\,\cite{Deut02}. For schemaless XML data,
we can derive a schema from XML data using the
DataGuide\,\cite{Gold97}\footnote{Recently, Bex et al.\,\cite{Bex07} have proposed algorithms for the inference of XML Schema Definitions, but we use the DataGuide since it takes linear time to create and has sufficient power for checking structural consistency. If a DTD or XML Schema are given along with XML data, we can exploit the given schema.}.
The DataGuide is a labeled tree that has every unique label path of
XML data. In a DataGuide, a node represents the label of an element\,(or
attribute), and an edge represents the parent-child relationship between
two nodes. A node in a DataGuide is uniquely identified by its label path. In this paper, we augment the DataGuide with keywords contained
in value nodes to support keyword queries at the schema level. We call
the augmented DataGuide {\it DataGuide}$^+$ and use it as the schema.
Every non-value node in a DataGuide$^+$ is assigned a preorder number\footnote{We can use other numbering schemes without loss of generality. For example, to handle schema evolution, we can use {\it Compact Dynamic Quaternary String}\,({\it CDQS}) encoding\,\cite{Li08a},
which allows for updates without the original nodes having to be renumbered.
}. Hereafter, we call a node of the DataGuide$^+$ a {\it schema node} to
distinguish it from a node of XML data, which we call an {\it instance
node}. For ease of explanation, we may refer to a schema node by its label path.

\begin{figure}[h]
\vspace{-0.5cm}
\centerline{\includegraphics[width=8.8cm]
{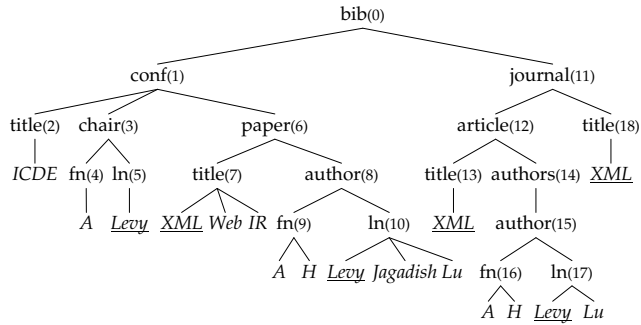}}
\caption{An example DataGuide$^+$.}
\label{fig:dataguide}
\ifx\vldbjformat\undefined \else \vspace{-0.6cm} \fi
\end{figure}

\begin{example}\label{eg:dataguide}
{\rm Fig.~\ref{fig:dataguide} shows the DataGuide$^+$ for the
XML data in Fig.~\ref{fig:motivating_example}(b). Every unique
label path of the XML data appears exactly once in the DataGuide$^+$.
For example, in the XML data, the label path
{\fssf ``bib.conf.paper.author"} appears twice, and so does
{\fssf ``bib.journal.article.authors.author"}. In contrast, in
the DataGuide$^+$, each appears only once.}\hfill$\Box$
\end{example}

\subsection{Query Models}
\subsubsection{Keyword Query}
We model a keyword query as a set of keywords\,\cite{Liu07}. As in the literature\,\cite{Bao09,Hris06,Hris03b,Huan08,Liu07,Liu08,Xu05}, each
query keyword may match (1) labels of elements or attributes or
(2) keywords contained in value nodes of the XML data.

\subsubsection{XPath Query}
We consider a subset of XPath that uses the child (``/'') and descendant (``//'') axes and predicates (``[]''). We model a query that belongs to this set as a twig pattern\,\cite{Brun02}. In the twig pattern a node, called a {\it query node}\,\cite{Brun02}, represents a label\,(or a value), and an edge represents the parent-child or ancestor-descendant relationship between
two nodes. One node of the twig pattern is marked as the {\it
query result node}\,\cite{Park05} and is distinguished from other
nodes by placing it in a box. A query node that has more than one
child node is called a {\it branching query node}\,\cite{Park05}. A leaf node of the twig pattern is called a {\it leaf
query node}.

\begin{figure}[h]
\centerline{\includegraphics[width=2cm]
{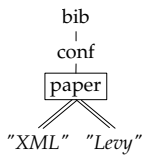}}
\caption{An example twig pattern.}
\label{fig:twig_pattern}
\end{figure}

\begin{example}
{\rm Fig.~\ref{fig:twig_pattern} shows an example twig pattern that
represents the XPath query $Q_1$. In Fig.~\ref{fig:twig_pattern},
{\fssf paper} is the query result node and, at the same time,
the branching query node. Keywords are located in leaf query nodes
{\fssf ``XML"} and {\fssf ``Levy"}.
}

\vspace{0.3cm}
$Q_1$: {\fssf /bib/conf/paper[``XML"][``Levy"]}\hfill$\Box$
\end{example}

\subsection{Quality Metrics of XML Keyword Search}
As quality metrics for
keyword queries, we use precision and recall, which have been widely
used in the field of information retrieval\,(IR). Formula
(\ref{eq:precision_recall}) shows the definitions of precision and
recall\,\cite{Baez99}. Here, $R$ is the set of  nodes relevant to the
query\,(i.e., desired results) in the database, and $A$ is the set of
nodes retrieved as the answer to the query\,(i.e., actual query
results). Precision is the fraction of the retrieved nodes\,(i.e.,
$A$) that are relevant, and recall is the fraction of the relevant
nodes\,(i.e., $R$) that have been retrieved. The search quality is
good when both precision and recall are close to 1.0\,\cite{Baez99}.

\vspace{-0.1cm}

\begin{eqnarray}
precision = \frac{|R \cap A|}{|A|},~recall = \frac{|R \cap
A|}{|R|}\label{eq:precision_recall}
\end{eqnarray} 
\section{Structural Consistency}
\label{sec:structural_consistency}

In this section, we formally define the notions of structural
consistency and structural anomaly in XML keyword search. We also
propose an efficient algorithm that resolves structural anomaly
at the schema level.

\subsection{The Concept}
\label{sec:concept}

We first define the {\it result structure} of a query result in
Definition~\ref{def:result_structure}. Here, a {\it query result} is a
subtree rooted at an SLCA in the XML data. We define {\it
structural containment} and {\it structural equivalence} of result structures in Definition~\ref{def:structural_comparison}. We then define the
{\it structural consistency} and the {\it structural anomaly} in
Definition~\ref{def:structural_consistency}.

\begin{definition}\label{def:result_structure}{\rm
The {\it result structure} of a query result $qr$, denoted as
$rs(qr)$, is a (schema-level) twig pattern composed of the label path from the root of XML data to the root $qr_{root}$ of $qr$\,(simply, the {\it
incoming label path}) and the ancestor-descendant edges from
$qr_{root}$ to query keywords. In the result structure $rs(qr)$, the node corresponding to $qr_{root}$ is marked as the query result node.}\hfill $\Box$
\end{definition}

In Definition~\ref{def:result_structure}, we note that the incoming label path information is sufficient to define the structural consistency, but we attach query keywords to find query results corresponding to the result structure in query processing.

\begin{figure}[h]
\ifx\vldbjformat\undefined \else \vspace{-0.4cm} \fi
\begin{center}
\begin{minipage}{4cm}
\begin{center}
\centerline{\includegraphics[width=3cm]
{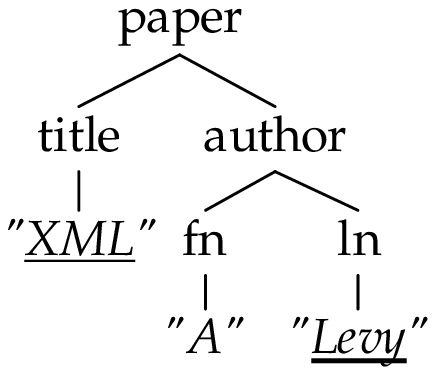}}
{\footnotesize (a) A query result {\sf paper(6)}}.
\end{center}
\end{minipage}
\begin{minipage}{4cm}
\begin{center}
\centerline{\includegraphics[width=3cm]
{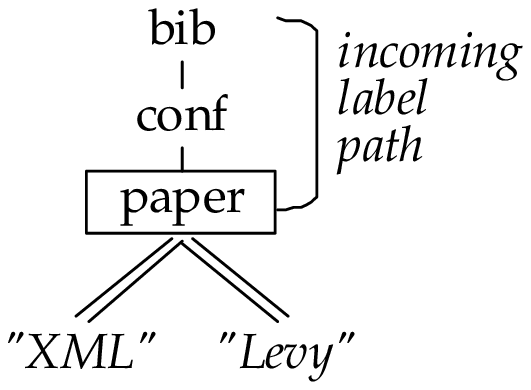}}
{\footnotesize (b) $rs$({\sf paper(6)})}.
\end{center}
\end{minipage}
\end{center}
\vspace{-0.2cm}
\caption{The result structure of a query result {\fssf paper(6)}.}
\label{fig:result_structure}
\vspace{-0.3cm}
\end{figure}

\begin{example}
{\rm Suppose that a keyword query $Q$ = {\fssf\{``XML",
``Levy"\}} is issued on the XML data in Fig.~\ref{fig:motivating_example}(a).
Fig.~\ref{fig:result_structure} shows a query result
{\fssf paper(6)} and its result structure. Note that a query result is a subtree of XML
data (i.e., an instance), and its result structure is a twig pattern (i.e., a part of schema).}\hfill $\Box$
\end{example}

\begin{definition}\label{def:structural_comparison}{\rm
Given a keyword query $Q$ and the set of query results $QR$\,=\,\{$qr_1$,\,$qr_2$,\,...,\,$qr_m$\} of $Q$, the result structure
$rs(qr_i)$ {\it structurally contains} the result structure $rs(qr_j)$, as denoted by $rs(qr_i)$\,$\prec$\,$rs(qr_j)$,
if and only if the incoming label path of $rs(qr_i)$ is a proper
prefix of that of $rs(qr_j)$.  $rs(qr_i)$ and $rs(qr_j)$ are {\it structurally equivalent}, as denoted by $rs(qr_i)$ $\equiv$ $rs(qr_j)$, if and
only if their incoming label paths are identical. We define
$rs(qr_i)$\,$\preceq$\,$rs(qr_j)$ as $rs(qr_i)$\,$\prec$\,$rs(qr_j)$ or
$rs(qr_i)$ $\equiv$ $rs(qr_j)$.}\hfill $\Box$
\end{definition}

\ifx\vldbjformat\undefined \else \vspace{0.1cm} \fi

\begin{definition}\label{def:structural_consistency}{\rm
Given a keyword query $Q$ and the set of query results $QR$\,=\,\{$qr_1$,\,$qr_2$,\,...,\,$qr_m$\} of $Q$, {\it structural consistency} is a property where the following condition is satisfied for $QR$: ($\forall$$qr_i$$\in$$QR$) (($\neg\exists$$qr_j$$\in$$QR$)($rs(qr_i)$\,$\prec$\,$ rs(qr_j)$)).
{\it Structural anomaly} is a property where structural consistency is violated, i.e., ($\exists$$qr_i$,\,$\exists$$qr_j$\,$\in$\,$QR$) ($rs(qr_i)$\,$\prec$\,$rs(qr_j)$).
}\hfill$\Box$
\end{definition}

\ifx\vldbjformat\undefined \else \vspace{0.1cm} \fi

\begin{figure}[h]
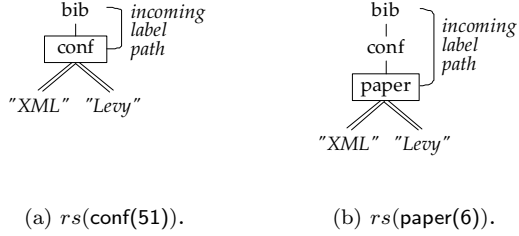

\vspace{-0.3cm}
\begin{center}
\begin{minipage}{4cm}
\begin{center}
\centerline{\includegraphics[width=2.8cm]
{figs/Structure_of_conf.eps}}
{\footnotesize (a) $rs$({\sf conf(51)})}.
\end{center}
\end{minipage}
\begin{minipage}{4cm}
\begin{center}
\centerline{\includegraphics[width=2.8cm]
{figs/Structure_of_paper.eps}}
{\footnotesize (b) $rs$({\sf paper(6)})}.
\end{center}
\end{minipage}
\end{center}
\vspace{-0.2cm}
\caption{The result structures of query results causing structural
anomaly.}
\label{fig:structural_anomaly}
\vspace{-0.1cm}
\end{figure}

\begin{example}
{\rm Suppose that a keyword query $Q$ = {\fssf\{``XML",
``Levy"\}} is issued on the XML data in Fig.~\ref{fig:motivating_example}(a), and that a set of query results $QR$ = {\fssf\{conf(51), paper(6)\}} is obtained. Fig.~\ref{fig:structural_anomaly} shows their result structures. We see that $rs$({\fssf conf(51)}) $\prec$ $rs$({\fssf paper(6)}). Thus, $QR$ has structural anomaly.}\hfill$\Box$
\end{example}

We resolve structural anomaly, thereby preserving structural consistency, by removing query results whose structure is not the
same as a {\it smallest result structure} as defined in
Definition~\ref{def:smallest_result_structure}.
By enforcing structural consistency, we can remove spurious results having the same result structure consistently.

\begin{definition}\label{def:smallest_result_structure}{\rm
Given a keyword query $Q$ and the set of query results $QR$ = \{$qr_1$, $qr_2$, ..., $qr_m$\} of $Q$, {\it the set of smallest
result structures} of $QR$ is \{$rs(qr_i)$\,$|$\,$qr_i$\,$\in$\,$QR$\,$\wedge$\,($\neg\exists$$qr_j$\,$\in$\,$QR$)
($rs(qr_i)$\,$\prec$\,$rs(qr_j)$)\}}~~~~$\Box$
\end{definition}

In Definition~\ref{def:smallest_result_structure}, ``smallest'' refers to the resulting subtrees since resulting subtrees are smaller if their incoming label paths are longer.

\begin{lemma}~\label{lemma:consistency}
{\rm Given a keyword query $Q$, the set of query results $QR$ =\,\{$qr_1$,\,$qr_2$,\,...,\,$qr_m$\} of $Q$, and the set of smallest result
structures $SRS$\,=\,\{$srs_1$,\,$srs_2$,\,...,\,$srs_n$\} of $QR$, structural consistency holds for $QR$ if the following
condition is satisfied for $QR$: ($\forall qr_i$\,$\in$\,$QR$)(($\exists
srs_j$\,$\in$\,$SRS$)($rs(qr_i)$\,$\equiv$\,$srs_j$)).}
\end{lemma}
\ifx\vldbjformat\undefined
\vspace{-0.4cm}
\else
\vspace{-0.2cm}
\fi
\noindent{\sc Proof:} It is straightforward from the definition of the smallest result structure.
\hfill$\Box$

\ifx\vldbjformat\undefined
\else
\vspace{0.4cm}
\fi

\begin{figure}[h]
\ifx\vldbjformat\undefined \else \vspace{-0.2cm} \fi
\centerline{\includegraphics[width=8.5cm]
{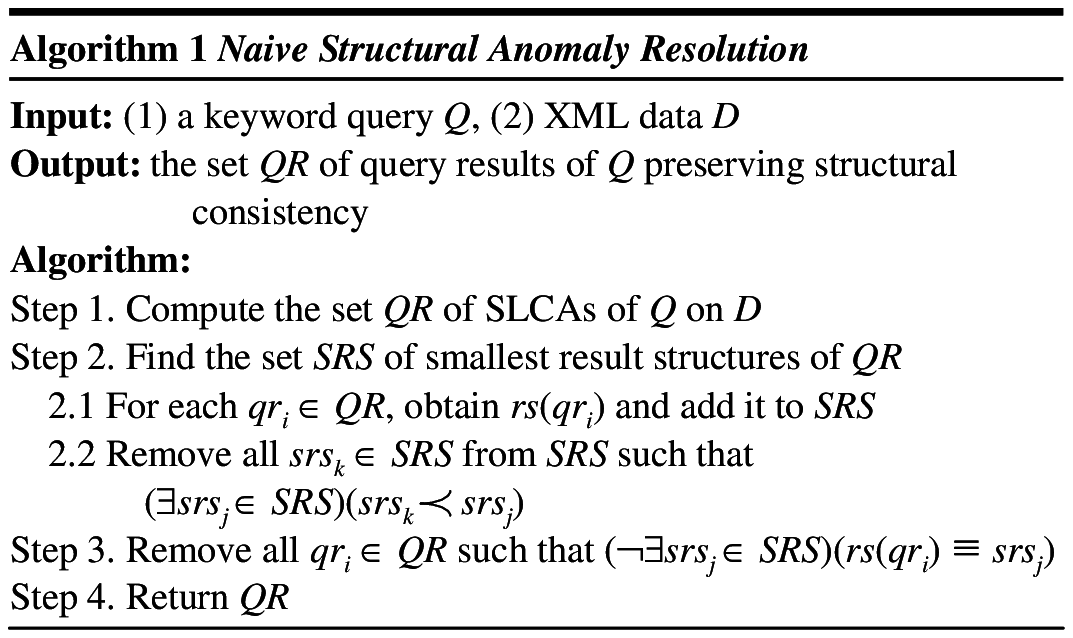}}
\ifx\vldbjformat\undefined \vspace{-0.6cm} \fi
\caption{A naive algorithm for resolving structural anomaly.}
\label{fig:Algorithm_Naive}
\end{figure}

Fig.~\ref{fig:Algorithm_Naive} shows a naive algorithm that resolves structural anomaly at the instance level. The algorithm consists of the following four steps: (1) computing all the SLCAs, (2) finding smallest result structures of the SLCAs, (3) removing SLCAs whose result structures are not smallest result structures, and (4) returning the set of SLCAs preserving structural consistency.

\subsection{Schema-level Structural Anomaly Resolution}
\label{sec:resolution}
Obviously, the naive algorithm would always be slower than existing
SLCA-based algorithms. We propose an efficient algorithm, called {\it
schema-level structural anomaly resolution}, that resolves structural
anomaly at the schema level. In this algorithm, we first find smallest result
structures at the schema level. We then compute only those query results
that correspond to the smallest result structures by evaluating structured queries constructed from the smallest result structures. We prove in Section~\ref{sec:schema_algorithm} that we can find the smallest result structures using the schema without incurring false dismissal. To do that we first define the {\it schema-level SLCA} in Section~\ref{sec:schema_slca}. We then formally analyze the relationship between the set of schema-level SLCAs and the set of instance-level SLCAs in Section~\ref{sec:relationship}. Through analysis, we show that simple query evaluation using the schema-level SLCAs cannot obtain the same query results as the instance-level algorithm does.
In Section~\ref{sec:generalization}, we present a solution for this problem, which we call {\it iterative $k$th-ancestor generalization}. In Section~\ref{sec:schema_algorithm}, we present a novel algorithm that resolves structural anomaly at the schema level using the schema-level SLCAs and iterative $k$th-ancestor generalization. We finally prove in Theorem~\ref{theorem:schema_level} that the schema-level algorithm and the
instance-level algorithm produce an equivalent set of query results that
preserve structural consistency.


\subsubsection{Schema-level SLCA}\label{sec:schema_slca}
We first define the {\it schema-level LCA} in Definition~\ref{def:slca} and then define the set of schema-level SLCAs in Definition~\ref{def:sslca}. In contrast, we call SLCAs in the XML data {\it instance-level SLCAs}.  Hereafter, {\it ancestor}($s_a$,\,$s$) denotes that node $s_a$ is an ancestor of node $s$, and {\it ancestor-or-self}($s_a$,\,$s$) denotes that {\it ancestor}($s_a$,\,$s$) or $s_a$\,=\,$s$.

\begin{definition}\label{def:slca}{\rm
Let $G$ be a DataGuide$^+$ and $S$ be the set of all schema nodes in
$G$. For $n$ schema nodes $s_1$,\,$s_2$,\,...,\,$s_n$\,$\in$\,$S$, $s_a$\,$\in$\,$S$ is the {\it schema-level LCA} of these $n$ schema nodes if and only if the following conditions are satisfied: (1) ($\forall$$1$$\leq$$i$$\leq$$n$) ({\it ancestor-or-self}($s_a$,\,$s_i$)), (2) ($\neg\exists s_b$\,$\in$\,$S$)({\it ancestor}($s_a$,\,$s_b$) $\wedge$ \\ ($\forall$$1$$\leq$$i$$\leq$$n$)({\it ancestor-or-self}($s_b$,\,$s_i$))). The schema-level LCA
$s_a$ for $s_1$,\,$s_2$,\,...,\,$s_n$ is denoted as $LCA$($s_1$,\,$s_2$,\,...,\,$s_n$).}\hfill $\Box$
\end{definition}

We note that, in Definition~\ref{def:slca}, the $LCA$ is defined for $n$
schema nodes; in Definition~\ref{def:sslca}, the $LCA\_SET$ is defined
for $m$ sets of schema nodes. Given a keyword query $Q$\,=\,\{$w_1$, $w_2$, ..., $w_m$\} and a DataGuide$^+$ $G$,  $S_i$ $(1$$\leq$$i$$\leq$$m)$
denotes the set of schema nodes directly containing $w_i$
in $G$.

\begin{definition}\label{def:sslca}{\rm
Given a keyword query $Q$\,=\,\{$w_1$, $w_2$, ..., $w_m$\} and the set
$S$ of all schema nodes in a DataGuide$^+$ $G$, {\it the set of schema-level SLCAs} $SLCA\_SET$($S_1$,\,$S_2$,\,...,\,$S_n$)\,=\,\{$s_a$\,$|$\,($s_a$ $\in$ $LCA\_SET$($S_1$,\,$S_2$,\,..., $S_n$))$\wedge$($\neg\exists$$s_b$\,$\in$\,$LCA\_SET$($S_1$,\,$S_2$,\,...,\,$S_n$)) ($ancestor$($s_a$,\,$s_b$))\} where $LCA\_SET$($S_1$,\,$S_2$,\,...,\,$S_m$) = \\ \{$s_a$ $|$ ($s_a$ $\in$ $S$) $\wedge$ ($\exists$ $s_1$ $\in$$S_1$, $\exists$ $s_2$ $\in$ $S_2$, ..., $\exists$ $s_m$ $\in$ $S_m$)($s_a$ = $LCA$($s_1$, $s_2$, ..., $s_m$)).}\hfill $\Box$
\end{definition}

\begin{example}
{\rm  Suppose that a keyword query $Q$ = {\fssf\{``XML", ``Levy"\}} is issued on the XML data in Fig.~\ref{fig:motivating_example}(b). In the DataGuide$^+$ in Fig.~\ref{fig:dataguide}, the set of schema-level LCAs is {\fssf\{``bib", ``bib.conf", ``bib.conf.paper", ``bib.journal", ``bib.journal.article"\}}, and the set of schema-level SLCAs is {\fssf\{``bib.conf.paper", ``bib.journal.article"\}} since
these schema nodes do not contain other schema-level LCAs.}\hfill
$\Box$
\end{example}

\subsubsection{The Relationship between
the Set of Schema-level SLCAs and the Set of Instance-level SLCAs}
\label{sec:relationship}

To explain the relationship between the set of schema-level SLCAs and
the set of instance-level SLCAs, we first define the {\it schema structure} of
a schema node in Definition~\ref{def:schema_structure}. Since both
the schema structure of a schema node and the result structure of a
query result are defined as twig patterns, we will use the same
notions of structural equivalence and structural containment for
schema structures.

\begin{definition}\label{def:schema_structure}{\rm
The {\it schema structure} of a schema node $s$, denoted as $ss(s)$,
is a twig pattern composed of the incoming label path from the root
of DataGuide$^+$ to $s$ and the ancestor-descendant edges from $s$
to query keywords. In the schema structure $ss(s)$, the node corresponding to  $s$ is marked as the query result node}.\hfill $\Box$
\end{definition}

\vspace{-0.3cm}

Given a keyword query, the set $SS$ of schema structures of schema-level
SLCAs is largely equivalent to the set $SRS$ of smallest result structures of instance-level SLCAs. However, there exist cases where $SS$ and $SRS$ are not equivalent since the schema loses some instance-level information by storing only unique label paths of the instance nodes. For example, in the XML data in
Fig.~\ref{fig:motivating_example}(a), {\fssf ``Levy"} and {\fssf ``Lu"} appear in the instance nodes with the label path {\fssf ``bib.conf.paper. author.ln"}, but they appear in different instance nodes, {\fssf ln(65)} and {\fssf ln(68)}. Nonetheless, in the DataGuide$^+$ in Fig.~\ref{fig:dataguide}, they appear in the same schema node with the label path {\fssf ``bib.conf. paper.author.ln"} since their label paths are the same. Thus, in effect, the schema loses the information that {\fssf ``Levy"} and {\fssf ``Lu"} appear in different
instance nodes with the same label path.

There are two cases where $SRS$ and $SS$ are not equivalent: case 1) for some $ss_j$\,$\in$\,$SS$, there exists an $srs_i$\,$\in$\,$SRS$ such that $srs_i$\,$\prec$\,$ss_j$, and case 2) for some $ss_j$\,$\in$\,$SS$, there exists no $srs_i$\,$\in$\,$SRS$ such that $srs_i$\,$\preceq$\,$ss_j$. We note that $ss_j$\,$\prec$\,$srs_i$ does not hold according to the definition of the schema-level SLCA. In case 1, if we compute query results corresponding to $ss_j$, we will miss query results corresponding to $srs_i$, i.e., we will incur {\it false dismissal}. Example~\ref{eg:Levy_Lu} shows an instance of false dismissal. In Section~\ref{sec:generalization}, we propose a solution to this problem, which we call {\it iterative $k$th-ancestor generalization}.
In case 2, if we blindly apply iterative $k$th-ancestor generalization for $ss_j$, we could end up with incurring structural anomaly. We call $ss_j$\,$\in$\,$SS$ such that ($\neg \exists srs_i$\,$\in$\,$SRS$)($srs_i$\,$\preceq$\,$ss_j$) a {\it phantom schema structure}. Example~\ref{eg:phantom} shows an example of the phantom schema structure. In the next section, we will provide a solution to eliminate phantom schema structures.

\vspace{-0.3cm}

\begin{figure}[h]
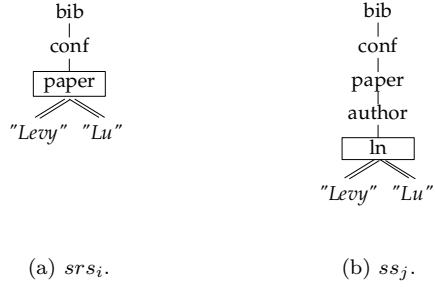

\vspace{-0.2cm}
\begin{center}
\begin{minipage}{4cm}
\begin{center}
\centerline{\includegraphics[width=1.7cm]
{figs/Levy_Lu_Structure2.eps}}
{\footnotesize (a) $srs_i$.}
\end{center}
\end{minipage}
\begin{minipage}{4cm}
\begin{center}
\centerline{\includegraphics[width=1.7cm]
{figs/Levy_Lu_Structure1.eps}}
{\footnotesize (b) $ss_j$.}
\end{center}
\end{minipage}
\end{center}
\vspace{-0.2cm}
\caption{An example of false dismissal.}
\label{fig:Levy_Lu_Structure2}
\vspace{-0.2cm}
\end{figure}

\begin{example}\label{eg:Levy_Lu}
{\rm Consider a keyword query $Q$ = \{{\fssf ``Levy"}, {\fssf ``Lu"}\} issued on the XML data in Fig.~\ref{fig:motivating_example}(a). Figs.~\ref{fig:Levy_Lu_Structure2}(a) and (b) show $srs_i$\,$\in$\,$SRS$ and
$ss_j$\,$\in$\,$SS$, respectively. Here, $srs_i$\,$\prec$\,$ss_j$. In the XML data in Fig.~\ref{fig:motivating_example}(a), we see that there is
a query result corresponding to $srs_i$, {\fssf paper(61)}, but
there is no query result corresponding to $ss_j$.}\hfill$\Box$
\end{example}

\begin{figure*}[hbt]
\begin{center}
\begin{minipage}{4.9cm}
\begin{center}
\centerline{\includegraphics[width=5.8cm]
{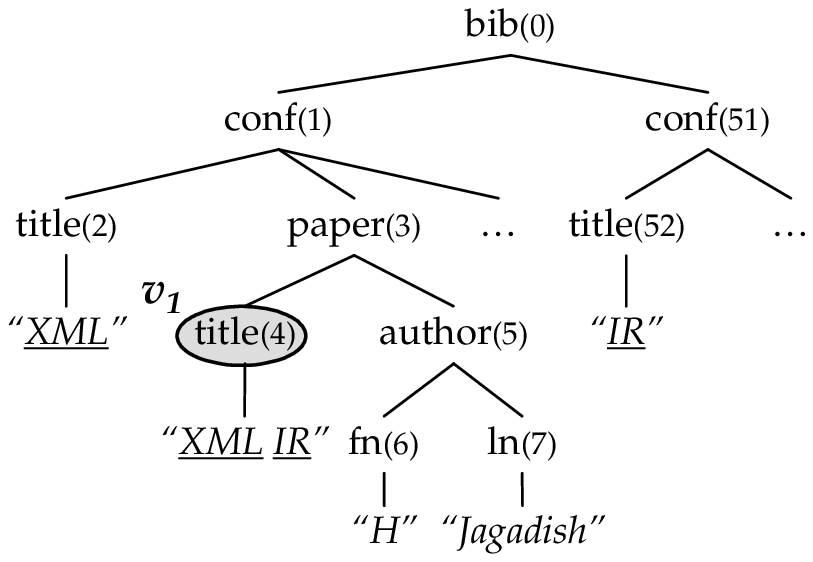}}
\ifx\vldbjformat\undefined \vspace{-0.2cm} \fi
{\footnotesize (a) XML data.}
\end{center}
\end{minipage}
\begin{minipage}{8cm}
\begin{center}
\centerline{\includegraphics[width=4.35cm]
{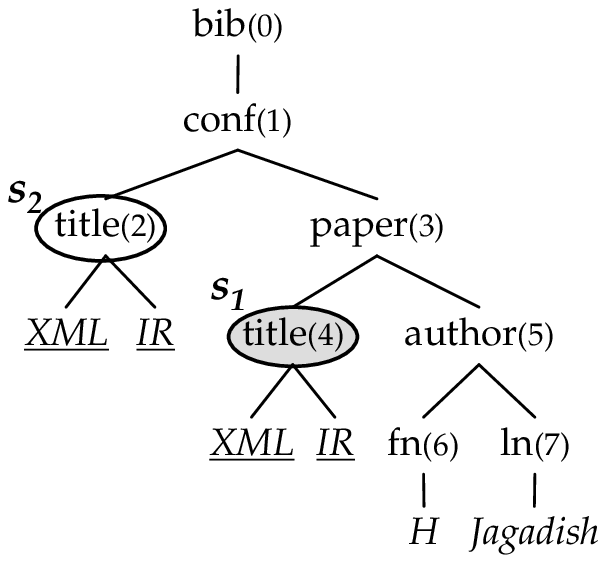}}
\ifx\vldbjformat\undefined \vspace{-0.2cm} \fi
{\footnotesize (b) The DataGuide$^+$ for the XML data in (a).}
\end{center}
\end{minipage}
\end{center}
\vspace{-0.2cm}
\caption{An example of a phantom schema structure.}
\label{fig:disjoint}
\vspace{-0.1cm}
\end{figure*}

\begin{example}\label{eg:phantom}
{\rm Suppose that a keyword query $Q$ = \{{\fssf ``XML"}, {\fssf ``IR"}\}
is used. In the XML data in Fig.~\ref{fig:disjoint}(a), $SRS$
= \{$rs(v_1)$\}. In the DataGuide$^+$ in
Fig.~\ref{fig:disjoint}(b), $SS$ = \{$ss(s_1), ss(s_2)$\}. Thus, we
do not have an $srs$ $rs(v_2)$ such that $rs(v_2)$\,$\preceq$\,$ss(s_2)$, and $ss(s_2)$ is a phantom schema structure. In this case, if we applied $k$th-ancestor generalization to $s_2$, we would find {\fssf conf(1)} in Fig.~\ref{fig:disjoint}(a) as a result, which causes structural anomaly because $rs$({\fssf conf(1)})\,$\prec$\,$rs$($v_1$). }\hfill$\Box$
\end{example}

We now formally state the relationship between $SRS$ and $SS$, which
will be used in iterative $k$th-ancestor generalization.

\begin{lemma}\label{lemma:preceq}
{\rm Given a keyword query $Q$, for all $srs_i$\,$\in$\,$SRS$, there exists
$ss_j$\,$\in$\,$SS$ such that $srs_i$\,$\preceq$\,$ss_j$.}
\end{lemma}
\ifx\vldbjformat\undefined
\vspace{-0.4cm}
\else
\vspace{-0.2cm}
\fi
\noindent{\sc Proof:} See Appendix A.\hfill$\Box$
\ifx\vldbjformat\undefined
\else
\vspace{0.3cm}
\fi

We can obtain $srs_i$\,$\in$\,$ SRS$ by computing the set $QR_j$ of the query results corresponding to $ss_j$\,$\in$\,$ SS$. If $QR_j$ is non-empty, then we have obtained $srs_i$\,$\in$\,$ SRS$ such that $srs_i$\,$\equiv$\,$ss_j$. If $QR_j$ is empty, we can obtain $srs_i$\,$\in$\,$SRS$ such that $srs_i$\,$\prec$\,$ss_j$ by applying iterative $k$th-ancestor generalization.

\subsubsection{Iterative $k$th-Ancestor
Generalization}\label{sec:generalization}

In this section, we present {\it iterative
$k$th-ancestor generalization} to solve the problems of false dismissal and phantom schema structures. Here, we iteratively find a $k$th-ancestor $s_a$ of the schema-level SLCA $s$ such that $ss(s_a)$\,$\equiv$\,$srs$\,$\in$\,$SRS$ where $srs \prec ss(s)$. We define the {\it $k$th-ancestor} in Definition~\ref{def:k_ancestor}.

\ifx\vldbjformat\undefined \vspace{-0.2cm} \fi

\begin{definition}\label{def:k_ancestor}{\rm
Given two nodes, $s_a$ and $s$, $s_a$ is the {\it $k$th-ancestor}
of $s$ if $s_a$ is an ancestor of $s$ and $depth(s)$\,=\,$depth(s_a)$\,+\,$k$ where $depth(s)$ is the length of the path from the
root to $s$.}\hfill $\Box$
\end{definition}

\ifx\vldbjformat\undefined \vspace{-0.4cm} \fi

\begin{example}\label{eg:generalization}
{\rm We can obtain $srs_i$\,$\in$\,$SRS$ in Fig.~\ref{fig:Levy_Lu_Structure2}(a) by finding the 2nd-ancestor of the schema-level SLCA in
Fig.~\ref{fig:Levy_Lu_Structure2}(b).}\hfill$\Box$
\end{example}

\ifx\vldbjformat\undefined \vspace{-0.4cm} \fi

\begin{lemma}\label{lemma:kth_ancestor}
{\rm Given a keyword query $Q$, suppose that $srs_i$\,$\in$\,$SRS$
structurally contains $ss(s)$\,$\in$\,$SS$, i.e., $srs_i$\,$\prec$\,$ss(s)$. Then, there
must exist a $k$th-ancestor $s_a$ (1\,$\leq$\,$k$\,$\leq$\,$depth(s)$) of $s$
such that $ss(s_a)$\,$\equiv$\,$srs_i$\,$\in$\,$SRS$.}
\end{lemma}
\ifx\vldbjformat\undefined
\vspace{-0.4cm}
\else
\vspace{-0.2cm}
\fi
\noindent{\sc Proof:} See Appendix B.\hfill $\Box$
\ifx\vldbjformat\undefined
\else
\vspace{0.3cm}
\fi

In iterative $k$th-ancestor generalization, we iteratively find the
$k$th-ancestor $s_a$ of the schema-level SLCA $s$ from the parent of
$s$ (i.e., $k$\,=\,1) until the set of the query results corresponding
to $ss(s_a)$ is non-empty. Here, obtaining non-empty results
indicates that $srs$\,$\in$\,$SRS$ has been found. Thus, we solve the false dismissal problem.

To eliminate phantom schema structures during iterative $k$th-ancestor generalization, we need to iteratively check structural consistency. Initially, there is no structural anomaly for the set of schema-level SLCAs. As schema-level SLCAs are generalized, structural anomaly can be incurred by their ancestors in the schema. Then, computing query results corresponding to the $k$th-ancestor incurring structural anomaly in the schema will incur structural anomaly in the instances. For example, in Fig.~\ref{fig:disjoint}(b), the schema structure of the 1st-ancestor of $s_2$, $ss$({\fssf conf(1)}), structurally contains the schema structure $ss(s_1)$ of the schema-level SLCA $s_1$. In this case, if we compute query results corresponding to $ss$({\fssf conf(1)}), we obtain {\fssf conf(1)} in Fig.~\ref{fig:disjoint}(a). Here, $rs$({\fssf conf(1)})\,$\prec$\,$rs$($v_1$) causing structural anomaly. Thus, we iteratively remove ancestors incurring structural anomaly and stop applying generalization for them. That is, we remove phantom schema structures.

We note that one $srs_i$\,$\in$\,$SRS$ can structurally contain multiple schema structures $ss(s_1)$, $ss(s_2)$, ..., $ss(s_n)$\,$\in$\,$SS$. In such cases, if we blindly generalize all the schema-level SLCAs $s_1$, $s_2$, ..., $s_n$, we obtain duplicate query results corresponding to $srs_i$. Thus, we must generalize only one schema-level SLCA for $srs_i$. This constraint is also enforced by iteratively checking structural consistency. Suppose that $s_1$, $s_2$, ..., $s_n$ are being generalized to $srs_i$ in this order. It is clear that $s_j$ (1$\leq$$j$$\leq$$n$-1) will be removed since $s_j$, when sufficiently generalized, must become the ancestor of $s_n$. Therefore, we can guarantee that only one schema-level SLCA, $s_n$, is generalized.

\subsubsection{Putting It Altogether}
\label{sec:schema_algorithm}

\begin{figure}[h]
\centerline{\includegraphics[width=8.5cm]
{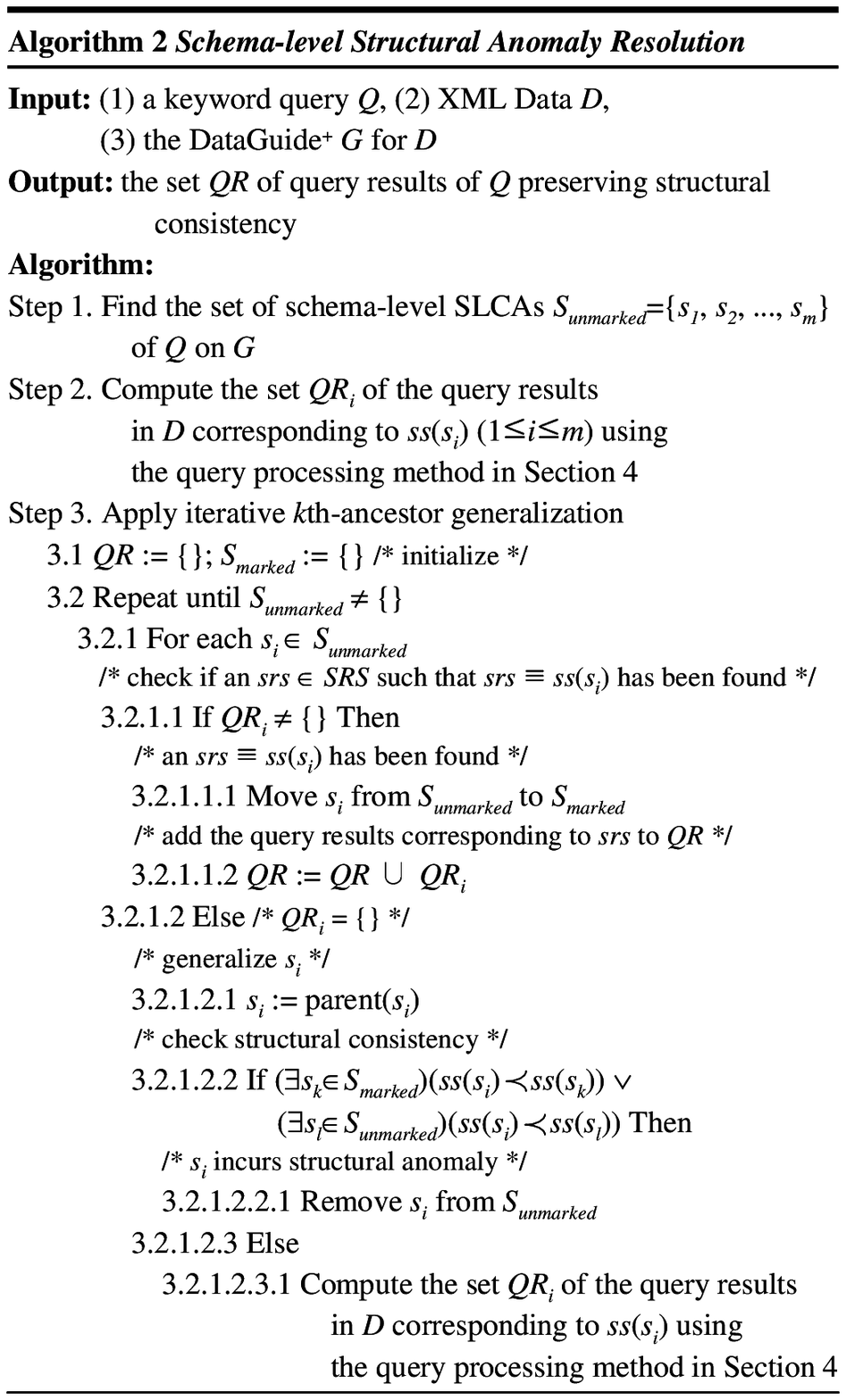}}
\caption{The algorithm for resolving structural anomaly at the schema-level.}
\label{fig:Algorithm_Schema_level}
\ifx\vldbjformat\undefined \else \vspace{-0.4cm} \fi
\end{figure}

Fig.~\ref{fig:Algorithm_Schema_level} shows an enhanced algorithm that resolves structural anomaly at the schema-level using the schema-level SLCAs and iterative $k$th-ancestor generalization. This algorithm produces the same query results as the instance-level algorithm in Fig.~\ref{fig:Algorithm_Naive} does. We will present the detailed query processing method of this algorithm in Section~\ref{sec:implementation}. Step 1 finds the set of schema-level SLCAs
$S_{unmarked}$\,=\,\{$s_1$, $s_2$, ..., $s_m$\}, and Step 2 computes the set of the query results corresponding to $ss(s_i)$ ($1$$\leq$$i$$\leq$$m$) by evaluating the XPath query that represent $ss(s_i)$.
Here, we convert $ss(s_i)$ to an XPath query to make our method run on top of {\it any} query evaluation engine that supports XPath.
Step 3 applies iterative $k$th-ancestor generalization for $s_i$\,$\in$\,$S_{unmarked}$. In Step 3.2.1.1, we check whether an $srs$\,$\in$\,$SRS$ such that $srs$\,$\equiv$\,$ss(s_i)$ has been found by examining whether $QR_i$ is non-empty. If it has, in Step 3.2.1.1.1, we move such $s_i$ to $S_{marked}$. If not, in Step 3.2.1.2.1, we obtain the parent of $s_i$ using the $parent(s_i)$ function. In Step 3.2.1.2.2.1, we remove $s_i$, which incurs structural anomaly, from $S_{unmarked}$.

\begin{example}
{\rm
Suppose that a keyword query $Q$ = \{{\fssf ``XML"}, {\fssf ``IR"}\}
is used to query the XML data in Fig.~\ref{fig:disjoint}(a).
In Step 1, $S_{unmarked}$ = \{$s_1$, $s_2$\}.
In Step 2, the set $QR_1$ of the query results corresponding to $ss(s_1)$ is non-empty (\{{\fssf title(4)}\}), but $QR_2$ for $ss(s_2)$ is empty.
In Step 3.2.1.1, since $QR_1 \neq \{\}$, we move $s_1$ from $S_{unmarked}$ to $S_{marked}$ and add $QR_1$ to the set $QR$ of query results. Hence, $S_{unmarked}$ = \{$s_2$\}, $S_{marked}$ = \{$s_1$\}, and $QR$ = \{{\fssf title(4)}\}. In Step 3.2.1.2, since $QR_2 = \{\}$, we generalize $s_2$. Now $s_2$ incurs structural anomaly since ($\exists s_1$\,$\in$\,$S_{marked}$)($ss(s_2)$\,$\prec$\,$ss(s_1))$. Thus, we remove $s_2$ from $S_{unmarked}$. Now $S_{unmarked} = \{\}$, and we end the iteration.

In Step 3, even if we process $s_2$ first, we can obtain the correct result without a problem. In Step 3.2.1.2.2, $s_2$ incurs structural anomaly since ($\exists s_1$\,$\in$\,$S_{unmarked}$)($ss(s_2)$\,$\prec$\,$ss(s_1))$. Thus, we remove $s_2$ from $S_{unmarked}$ obtaining $S_{unmarked}$ = \{$s_1$\} and $S_{marked}$ = \{\}. Now we move $s_1$ from $S_{unmarked}$ to $S_{marked}$, add $QR_1$ to $QR$, and end the iteration.\hfill$\Box$}
\end{example}

\ifx\vldbjformat\undefined \else \vspace{0.1cm} \fi
\begin{theorem}\label{theorem:schema_level}
{\rm The Schema-level Structural Anomaly Resolution algorithm produces the same query results as the instance-level algorithm in Fig.~\ref{fig:Algorithm_Naive} does.}
\end{theorem}
\ifx\vldbjformat\undefined \vspace{-0.4cm} \else \vspace{-0.2cm} \fi
\noindent{\sc Proof:}
By Lemma~\ref{lemma:preceq}, for every
$srs_i$\,$\in$\,$SRS$, there exists $ss(s_j)$ $\in$ $SS$ such that (1) $srs_i$\,$\equiv$\,$ss(s_j)$ or (2) $srs_i$\,$\prec$\,$ss(s_j)$. For case 1, we can obtain $srs_i$\,$\in$\,$SRS$ by computing the query results corresponding to $ss(s_j)$ (Step 2). For case 2, we can obtain $srs_i$\,$\in$\,$SRS$ by applying iterative $k$th-ancestor generalization according to Lemma~\ref{lemma:kth_ancestor} (Step 3). In this case, even if generalization is stopped for $s_j$ because of incurring structural anomaly, we are still able to obtain $srs_i$\,$\in$\,$SRS$ since there always exists a schema-level SLCA $s_n$ such that $ss(s_j)$\,$\prec$\,$ss(s_n)$ ---which is exactly what caused the structural anomaly---and we can find $srs_i$ by generalizing $s_n$.
Finally, $ss(s_j)$\,$\in$\,$SS$ such that ($\neg \exists srs_i$\,$\in$\,$ SRS$)($srs_i$\,$\preceq$\,$ss(s_j)$), i.e., the phantom schema structure, is always removed since the $k$th-ancestor $s_a$ of $s_j$ must eventually incur structural anomaly when $s_j$ is generalized to the root node. Otherwise, we contradict the assumption ($\neg \exists srs_i$\,$\in$\,$SRS$)($srs_i$\,$\preceq$\,$ss(s_j)$) since it must be that $srs_i$\,$\equiv$\,$ss(s_a)$ at the root node.\hfill$\Box$
\ifx\vldbjformat\undefined \else \vspace{0.3cm} \fi

We now analyze the complexity of our schema-level algorithm.
Given a keyword query $Q$\,=\,\{$w_1$,\,$w_2$, ..., $w_n$\}, the worst case time complexity of the schema-level algorithm is $O(|S_1| d$ $\sum^n_{i=2} log |S_i| + d C_{XPath})$ where $S_i$ $(1$$\leq$$i$$\leq$$n)$ is the set of schema  nodes directly containing the query keyword $w_i$ in the DataGuide+, $d$ the maximum depth of the XML data, and $C_{XPath}$ the cost of XPath query evaluation, which will be presented in Section~\ref{sec:query_evaluation}. Here, $O(|S_1| d \sum^n_{i=2} log |S_i|)$\,\cite{Xu05} is the cost of computing schema-level SLCAs using the algorithm of Xu and Papakonstantinou\,\cite{Xu05}, and $O(d C_{XPath})$ is the cost of iterative $k$th-ancestor generalization since, in the worst case, generalization can be applied
until one of the schema-level SLCAs reaches the root node.

Compared with the existing instance-level SLCA algorithm\,\cite{Xu05}, the schema-level algorithm is generally more efficient since it avoids unnecessary computation of spurious results by removing them early at the schema-level. The additional overheads of the schema-level algorithm are the computation of schema-level SLCAs and iterative $k$th-ancestor generalization. However, those overheads are small in practice. First, the cost of the schema-level SLCA computation tends to be very small since the schema is generally several orders of magnitude smaller than the XML data\,\cite{Ario08}. Second, the cost of iterative $k$th-ancestor generalization is negligible since the generalization occurs only occasionally and is usually applied only once or twice. (According to our experiments in Section~\ref{sec:experiments}, the cost of iterative $k$th-ancestor generalization is less than 10\% of the total query processing cost.) In the worst case, however, our schema-level algorithm could be about twice slower than the instance-level SLCA algorithm. The reasons are as follows. First, when the schema is as large as the XML data, the overhead of schema-level SLCA computation would be almost the same as the cost of the instance-level SLCA computation. Second, after obtaining the schema-level SLCAs, we compute query results that correspond to the schema-level SLCAs by evaluating the XPath queries. This query evaluation could also be as expensive as the instance-level SLCA computation if there exist few spurious results since then our method loses the benefit over existing SLCA-based methods of avoiding unnecessary computation of spurious results through early removal. (See the experimental results of $QD_1$ and $QD_5$ in Fig.~\ref{fig:dblp_result1}(c) and $QX_1$ and $QX_8$ in Fig.~\ref{fig:xmark_result1}(c) of Section~\ref{sec:experiments}.)

\vspace{-0.5cm}

\subsection{A Relevance-Feedback Based Solution for the Low Recall Problem}
\label{sec:feedback}

\vspace{-0.3cm}

When users intend to find more general results (although this is relatively rare), which we regard as spurious results, our method can have lower recall than existing methods. For example, suppose that a user intends to find a conference on {\fssf``XML"} where {\fssf``Levy"} is the chair. If there is at least one paper about {\fssf``XML"} authored by {\fssf``Levy"}, our method does not retrieve the desired conference. We call this problem the {\it low recall problem}.

The fundamental cause for this problem is the inherent ambiguity in keyword search, i.e., the actual intention of the user is unknown. We can solve this problem by exploiting the user's relevance feedback. Relevance feedback is an important way of enhancing search quality by using relevance information provided by the user\,\cite{Hlao07,Sche06}. The solution is as follows. The initial query results are presented to the user, and the user gives feedback if desired results are not retrieved. (This kind of relevance feedback can be easily implemented using a user-friendly GUI, and users just need to click a button.) This feedback is sent to the system, and the system generalizes the smallest result structure and finds results again.
(We can repeat this feedback process until all the desired results are retrieved.) For example, our method does not retrieve the desired conference if there is at least one paper about {\fssf``XML"} authored by {\fssf``Levy"}. Since the desired result has not been retrieved, the user sends feedback to the system, and the system now finds conferences containing {\fssf``XML"} and {\fssf``Levy"} by generalizing the smallest result structure. Then, the user can obtain the desired result. When there are multiple smallest result structures, we can allow the user to choose which smallest result structure he wants to generalize. To do this, we need to group the query results for each smallest result structure and show each group to the user.

We implement this relevance-feedback based solution by modifying Algorithm 2. In Step 3.2.1.1 of Algorithm 2, we check whether the set $QR_i$ of the query results corresponding to a schema-level SLCA $s_i$ is non-empty. If $QR_i$ is empty, we generalize $s_i$ in Step 3.2.1.2.1 by finding the parent of $s_i$. We implement relevance feedback by modifying Step 3.2.1.1 such that $s_i$ should be generalized even if $QR_i$ is non-empty when the user's relevance feedback is received.

The reason why relevance feedback is possible is that we process queries at the schema level. The schema-level processing makes the relevance-feedback mechanism feasible since users just need to give feedback on a small number of schema-level SLCAs. However, it is hard to apply to instance-level methods since the number of instance-level SLCAs is generally much larger than that of schema-level SLCAs. Furthermore, it is not clear how we can receive the relevance feedback and generalize the results in the instance-level SLCA algorithm\,\cite{Xu05}.

\begin{figure}[h!]
\ifx\vldbjformat\undefined \else \vspace{-0.6cm} \fi
\centerline{\includegraphics[width=2.8cm]
{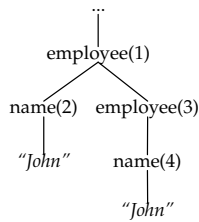}}
\ifx\vldbjformat\undefined \vspace{-0.8cm} \else \fi
\caption{XML data having a recursive schema.}
\label{fig:recursive}
\ifx\vldbjformat\undefined \else \vspace{-0.4cm} \fi
\end{figure}

We can handle XML data having a recursive schema using the same technique. Fig.~\ref{fig:recursive} shows recursive XML data where the parent-child relationship between two employees represents the supervisor-supervisee relationship. Suppose that the query is {\fssf``John employee''} and the user intends to find all employees whose name is {\fssf``John''}.
In this case, our method (and also SLCA and MLCA) finds only {\fssf employee(3)}, resulting in low recall. We can also resolve this problem by generalizing the smallest result structure via relevance feedback.

The low recall problem may also be handled by ranking in a spirit similar to the work of Amer-Yahia et al.\,\cite{Amer05}. Enabling users to exploit partial knowledge of the schema in user queries\,\cite{Cohe03,Li08,Yu07} can also help us to disambiguate user's intention. We leave these issues for future work.

\subsection{Search Quality Comparisons with Earlier Methods}
\label{sec:comparision}
In this section, we summarize search quality comparisons with earlier methods, SLCA\,\cite{Xu05}, MLCA\,\cite{Li08} (a variant of SLCA), XSEarch\,\cite{Cohe03}, CVLCA\,\cite{Li07}, and XReal\,\cite{Bao09}.
XSEarch and CVLCA are based on a heuristic called {\it interconnection} relationship. According to the heuristic, two nodes are considered to be semantically related if and only if there are no two distinct nodes with the same label on the path between these two nodes (excluding the two nodes themselves). Li et al.\,\cite{Li08} have pointed out that the heuristic could retrieve spurious results and have shown that MLCA is generally superior to the heuristic. XReal infers the user's intention using the statistics of the underlying XML data.

Since keyword queries are inherently ambiguous, the desired results of a keyword query depend on the user's intention. The user may want to find 1) more specific results or 2) more general (as opposed to specific) results. For example, for a keyword query {\fssf``XML Levy"}, the user may want to find either 1) papers about {\fssf``XML"} authored by {\fssf``Levy"} or 2) conferences on {\fssf``XML"} where {\fssf``Levy"} is the chair.

When the user's intention is to find more specific results, the precision values of our method are higher than or equal to those of existing methods since our method is able to eliminate more spurious results (i.e., general results) than existing methods by enforcing structural consistency. In addition, the recall values of our method and those of existing methods are the same since our method finds all the specific results, i.e., the query results that correspond to smallest result structures, as existing methods do.

\begin{figure*}[hbt]
\ifx\vldbjformat\undefined \vspace{-0.6cm} \else \fi
\centerline{\includegraphics[width=14cm]
{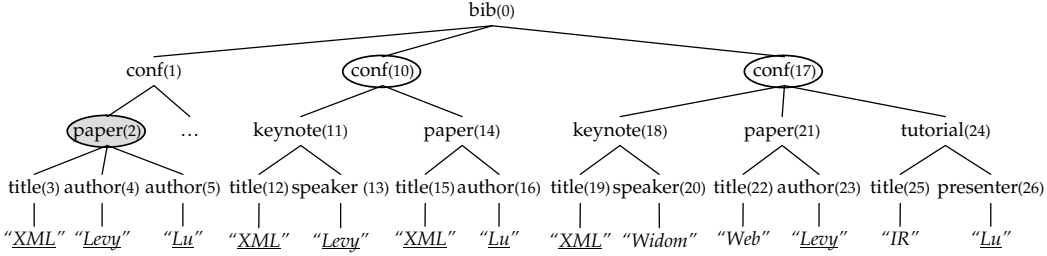}}
\ifx\vldbjformat\undefined \vspace{-0.6cm} \else \fi
\caption{The case where structural consistency shows high precision.}
\label{fig:XML_Levy_Lu}
\end{figure*}

\ifx\vldbjformat\undefined \vspace{-0.4cm} \else \fi
\begin{example}
{\rm Suppose that a keyword query $Q$ = \{{\fssf ``XML"}, {\fssf ``Levy"}, {\fssf ``Lu"}\} is issued on the XML data in Fig.~\ref{fig:XML_Levy_Lu}. The user wants to find papers about {\fssf``XML"} authored by {\fssf``Levy"} and {\fssf``Lu"}, and the desired result is {\fssf paper(2)}. SLCA, XSEarch, and CVLCA find not only {\fssf paper(2)} but also spurious (i.e., general) results {\fssf conf(10)} and {\fssf conf(17)}. MLCA can eliminate {\fssf conf(10)} since in the subtree rooted at {\fssf conf(10)}, {\fssf title(12)} and {\fssf title(15)} are the nodes that contain {\fssf``XML"}, and {\fssf speaker(13)} is the node that contains {\fssf ``Levy"} and the LCA of {\fssf title(15)} and {\fssf speaker(13)}, i.e., {\fssf conf(10)}, contains the LCA of {\fssf title(12)} and {\fssf speaker(13)}, i.e., {\fssf keynote(11)}. XReal retrieves \{{\fssf conf(10)}, {\fssf conf(17)}\} with the ranking since it infers {\fssf conf} as the desired node type\footnote{Since the highest confidence value (2.66) is significantly higher than the second highest value (1.41), XReal chooses the one with the highest confidence, {\fssf conf}, as the desired node type and retrieves only {\fssf conf} nodes.} based on the XML document frequency\,\cite{Bao09}. Our method can eliminate all the spurious results by enforcing structural consistency. Thus, compared with SLCA, MLCA, XSEarch, CVLCA, and XReal, our method improves precision without sacrificing recall.}\hfill$\Box$
\end{example}


\begin{figure}[h!] 
\vspace{-0.3cm}
\centerline{\includegraphics[width=15cm]
{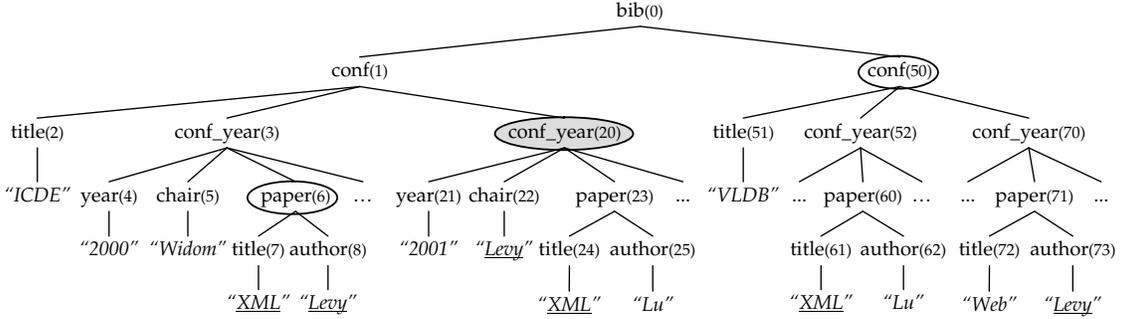}}
\ifx\vldbjformat\undefined \vspace{-0.4cm} \else \fi
\caption{The case where structural consistency shows low recall.}
\label{fig:XML_Levy_Conf}
\end{figure} 

When the user's intention is to find more general results, our method can have lower recall than existing methods, and we can solve this problem using relevance feedback. The recall values of our method with relevance feedback are higher than or equal to those of existing methods since we can eventually obtain the desired results via generalization. In the worst case, however, the precision values of our method with relevance feedback could be lower than those of existing methods since it may find more spurious results during generalization as we see in Example~\ref{eg:low_recall}. We note that the worst case is quite rare in practice.\footnote{To find one, we had to test more than one hundred queries that are structurally similar to that shown in Example~\ref{eg:low_recall} against the NASA and XMark data sets in Section~\ref{sec:experiments}. We were not able to find a similar query in the DBLP data set since its structure is simpler than those of the NASA and XMark data sets.}

\ifx\vldbjformat\undefined \vspace{-0.4cm} \else \vspace{0.2cm} \fi
\begin{example}\label{eg:low_recall}
{\rm
Suppose that a keyword query $Q$ = \{{\fssf ``XML"}, {\fssf ``Levy"}\} is issued on the XML data in Fig.~\ref{fig:XML_Levy_Conf} to find conferences on {\fssf``XML"} where {\fssf``Levy"} is the chair. The desired result is {\fssf conf\_year(20)}. SLCA and MLCA find \{{\fssf paper(6)}, {\fssf conf\_year(20)}, {\fssf conf(50)}\}. XSEarch and CVLCA find \{{\fssf paper(6)}, {\fssf conf\_year(20)}\}. XReal finds \{{\fssf conf\_year(3)}, {\fssf conf\_year(20)}\}. Here, {\fssf paper(6)}, \ifx\vldbjformat\undefined {\fssf conf\_year(3)} \else {\fssf conf\_} {\fssf year(3)}\fi, and {\fssf conf(50)} are spurious results. Our method initially finds only \{{\fssf paper(6)}\}, and thus, the recall of our method is 0. By using relevance feedback, our method obtains \{{\fssf conf\_year(3)}, {\fssf conf\_year(20)}\} through generalization, and thus, the recall becomes 1.0.
During generalization, our method finds a spurious result {\fssf conf\_year(3)}, but the precision value of our method is higher than those of SLCA and MLCA since the subtree rooted at {\fssf conf(50)} is much bigger than that of {\fssf conf\_year(3)}. However, if we remove the subtree rooted at {\fssf conf(50)} from the XML data (this is the worst case of our method), the precision value of our method can be lower than those of SLCA and MLCA. (See Figs.~\ref{fig:QN5}(a) and \ref{fig:QX5}(a) in Section~\ref{sec:exp_results}.) Compared with XSEarch and CVLCA, the precision value of our method is lower since our method finds {\fssf conf\_year(3)}. Compared with XReal, the precision value of our method is lower since our method  finds {\fssf paper(6)}.}\hfill$\Box$
\end{example}

\section{Implementation}
\label{sec:implementation}
In this section, we describe the implementation details of the schema-level structural anomaly resolution. Section~\ref{sec:index_structure} presents the index structures used in the query processing. Section~\ref{sec:query_processing_method} presents the query processing method.
\subsection{Index Structures}\label{sec:index_structure}
To speed up query processing, we use indexes for the Data-Guide$^+$ and XML data. We use an inverted index for a Data-Guide$^+$, which we call the {\it schema index}, to efficiently compute the schema-level SLCAs. We use an inverted index for XML data, which we call the {\it instance index}, to efficiently evaluate XPath queries. Inverted indexes have
been used in many XML query processing methods\,\cite{Brun02,Guo03,Li08,Park05}. We also use a table called {\it LabelPath}\,\cite{Park05} to store all the label paths occurring in the DataGuide$^+$.

Table~\ref{tbl:notation} summarizes the notation to be used for
explaining the index structures. In Table~\ref{tbl:notation}, if a schema (or an instance) node $s$ is a value node, we use $parent(s)$ instead of $s$ as a parameter for all functions since value nodes themselves do not have ids.

\ifx\vldbjformat\undefined \renewcommand{\baselinestretch}{1.0} \fi

\begin{table}
\begin{center}
\caption{Summary of notation.}
\ifx\vldbjformat\undefined \vspace{0.1cm} \fi
\label{tbl:notation}
{\small
\begin{tabular}{|c|l|}
\hline Symbols & \multicolumn{1}{c|}{Definitions}
\\ \hline \hline $snode\_id(s)$ & the id of a schema node $s$
\\ \hline $label\_path(s)$ & the label path of a schema \\ & (or an instance) node $s$
\\ \hline $label\_path\_id(s)$ & the id of $label\_path(s)$ =
$snode\_id(s)$
\\ \hline  & $label\_path(s)$ represented as a sequence \\ $numeric\_label\_path(s)$ & of $snode\_id$s rather than labels \\ & ($numeric\_label\_path(s)[i]$ denotes \\ & the $i$th id.)
\\ \hline $inode\_id(o)$ & the id of an instance node $o$
\\ \hline $node\_path(o)$ & the node path of an instance node $o$
\\ \hline
\end{tabular}
}
\end{center}
\ifx\vldbjformat\undefined
\end{table} \renewcommand{\baselinestretch}{2.0}
\else
\vspace*{-0.4cm} \end{table}
\fi

A LabelPath table consists of tuples of the form $\langle$\ifx\vldbjformat\undefined$label\_path\_id$\else$label\_$ $path\_id$\fi, $label\_path$$\rangle$, where
$label\_path$ is the label path of a schema node $s$, and $label\_path\_id$ is the same as the id of $s$. A B$^+$-tree index is created on the $label\_path\_id$ column, and an inverted index on the $label\_path$ column.

\begin{figure}[h]
\ifx\vldbjformat\undefined \vspace{-0.2cm} \else \vspace{-0.8cm} \fi
\centerline{\includegraphics[width=6cm]
{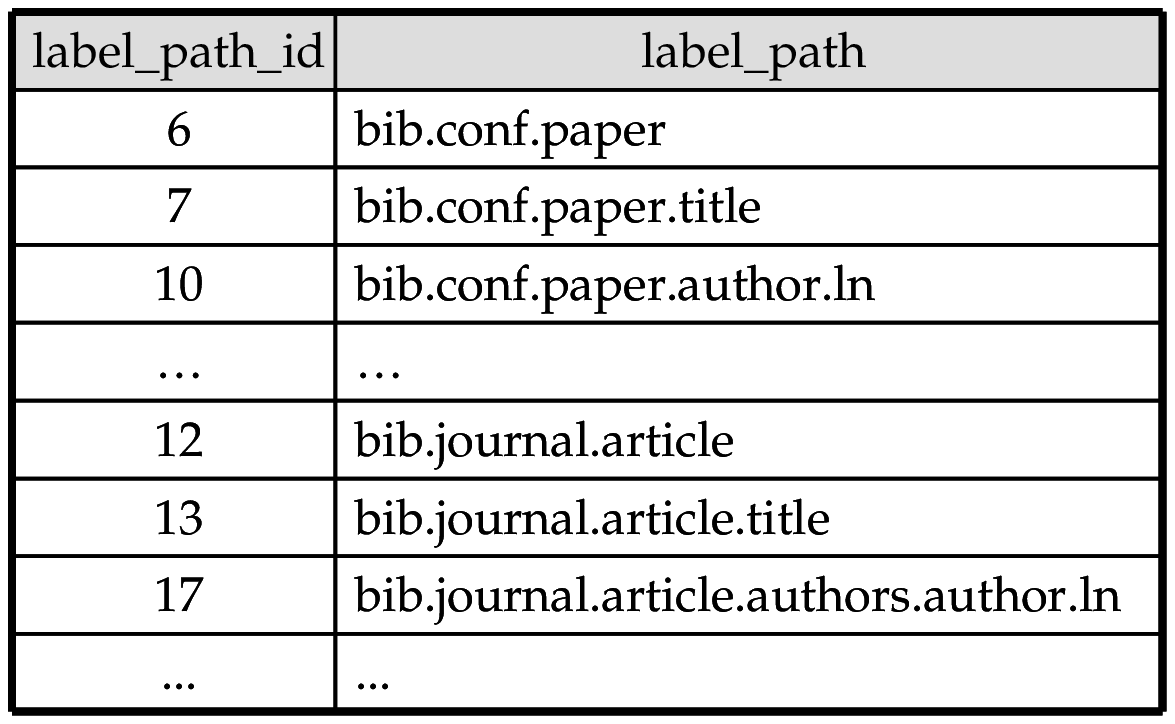}}
\ifx\vldbjformat\undefined \vspace{-0.2cm} \fi
\caption{An example LabelPath table.}
\label{fig:labelpath_table}
\end{figure}

\begin{example}
{\rm Fig.~\ref{fig:labelpath_table} shows the LabelPath table for the
\ifx\vldbjformat\undefined DataGuide$^+$\else Data-Guide$^+$\fi in Fig.~\ref{fig:dataguide}. In the DataGuide$^+$, the label path of the schema node having the id of {\fssf 6} is {\fssf ``bib.conf.paper"}.}\hfill$\Box$
\end{example}

The schema index stores a list of postings for each unique
value\,(or label) that appears in the DataGuide$^+$. The posting of
a schema node $s$ has the form $\langle$$snode\_id(s)$, \ifx\vldbjformat\undefined$numeric\_label\_path(s)$\else$numeric\_label\_$ $path(s)$\fi$\rangle$. $numeric\_label\_path(s)$ is used to find the ancestor nodes of $s$. Postings in a posting list are stored in ascending order of $snode\_id(s)$.

\begin{figure}[h]
\vspace*{-0.40cm}
\centerline{\includegraphics[width=8.5cm]
{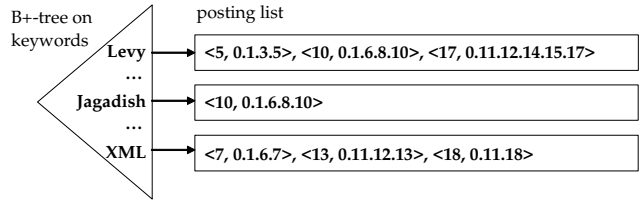}}
\ifx\vldbjformat\undefined \vspace{-0.3cm} \fi
\caption{An example schema index.}
\label{fig:schema_index}
\vspace*{-0.20cm}
\end{figure}

\vspace*{0.20cm}
\begin{example}
{\rm Fig.~\ref{fig:schema_index} shows the schema index for the
Data-Guide$^+$ in Fig.~\ref{fig:dataguide}. Let $s$ be the schema
node with the value = {\fssf ``Jagadish"} in
Fig.~\ref{fig:dataguide}. Then, $snode\_id(s)$ = {\fssf 10} and \ifx\vldbjformat\undefined$numeric\_label\_path(s)$\else$numeric\_label\_$ $path(s)$\fi = {\fssf 0.1.6.8.10}. Thus, a posting $\langle${\fssf 10}, {\fssf 0.1.6.8.10}$\rangle$ is stored in the posting list of {\fssf ``Jagadish"}.}\hfill$\Box$
\end{example}

The instance index stores a list of postings for each unique keyword\,(or label) that appears in XML data. The posting of an instance node $o$ has the form $\langle$$inode\_id(o)$, $node\_path(o)$, $numeric\_label\_path(o)$$\rangle$. $node\_path(o)$ is used to find the ancestor nodes of $o$, and $numeric\_label\_path(o)$ is used to find the label path of $o$. Postings in a posting list are stored in ascending order of $ inode\_id(o)$. We create a B$^+$-tree index, which is called a {\it subindex}\,\cite{Wha02,Whan05}, on each posting list of the instance index in the same way as was done by Guo et al.\,\cite{Guo03} and Whang et al.\,\cite{Wha02,Whan05}. The key of a subindex is $inode\_id(o)$.

\begin{figure}[h] 
\centerline{\includegraphics[width=11.5cm]{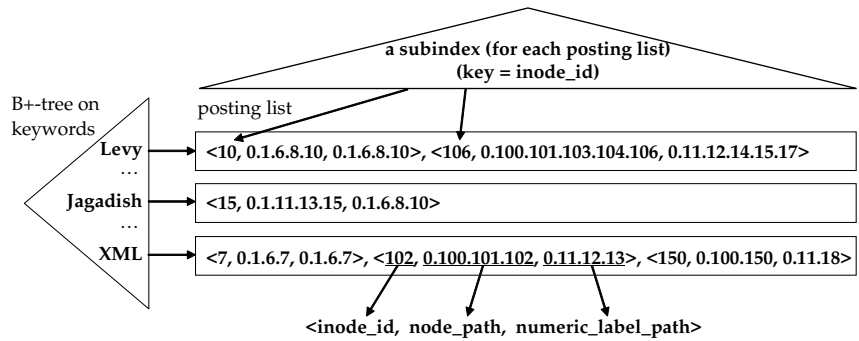}}
\ifx\vldbjformat\undefined \vspace*{-0.60cm} \fi
\caption{An example instance index.}
\label{fig:instance_index}
\vspace*{-0.40cm}
\end{figure}

\begin{example}
{\rm Fig.~\ref{fig:instance_index} shows the instance index for the XML data in Fig.~\ref{fig:motivating_example}(b). Let $o$ be the instance node with the value = {\fssf ``Jagadish"} in Fig.~\ref{fig:motivating_example}(b). Then, $inode\_id(o)$ = {\fssf 15}, $node\_path(o)$ = {\fssf 0.1.11.13.15}, and $label\_path(o)$ = {\fssf ``bib.conf.paper.author.ln"}. Since $numeric\_label\_path(o)$ = {\fssf 0.1.6.8.10} for $label\_path(o)$ in the Data-Guide$^+$ in Fig.~\ref{fig:dataguide}, a posting $\langle${\fssf
15}, {\fssf 0.1.11.13.15}, \ifx\vldbjformat\undefined \else \\ \fi{\fssf 0.1.6.8.10}$\rangle$ is stored in the posting list of {\fssf ``Jagadish"}.}
\hfill $\Box$
\end{example}

\subsection{Query Processing Method}
\label{sec:query_processing_method}
\vspace*{-0.2cm} 

The query processing method consists of the following two steps. The first step presented in Section~\ref{sec:query_translation} translates a given keyword query $Q$ into multiple XPath queries
corresponding to the schema-level SLCAs. The second step presented in Section~\ref{sec:query_evaluation} evaluates the
XPath queries obtained in the first step.

\subsubsection{Query Translation}
\label{sec:query_translation}

\begin{figure}[h]
\centerline{\includegraphics[width=8.5cm]
{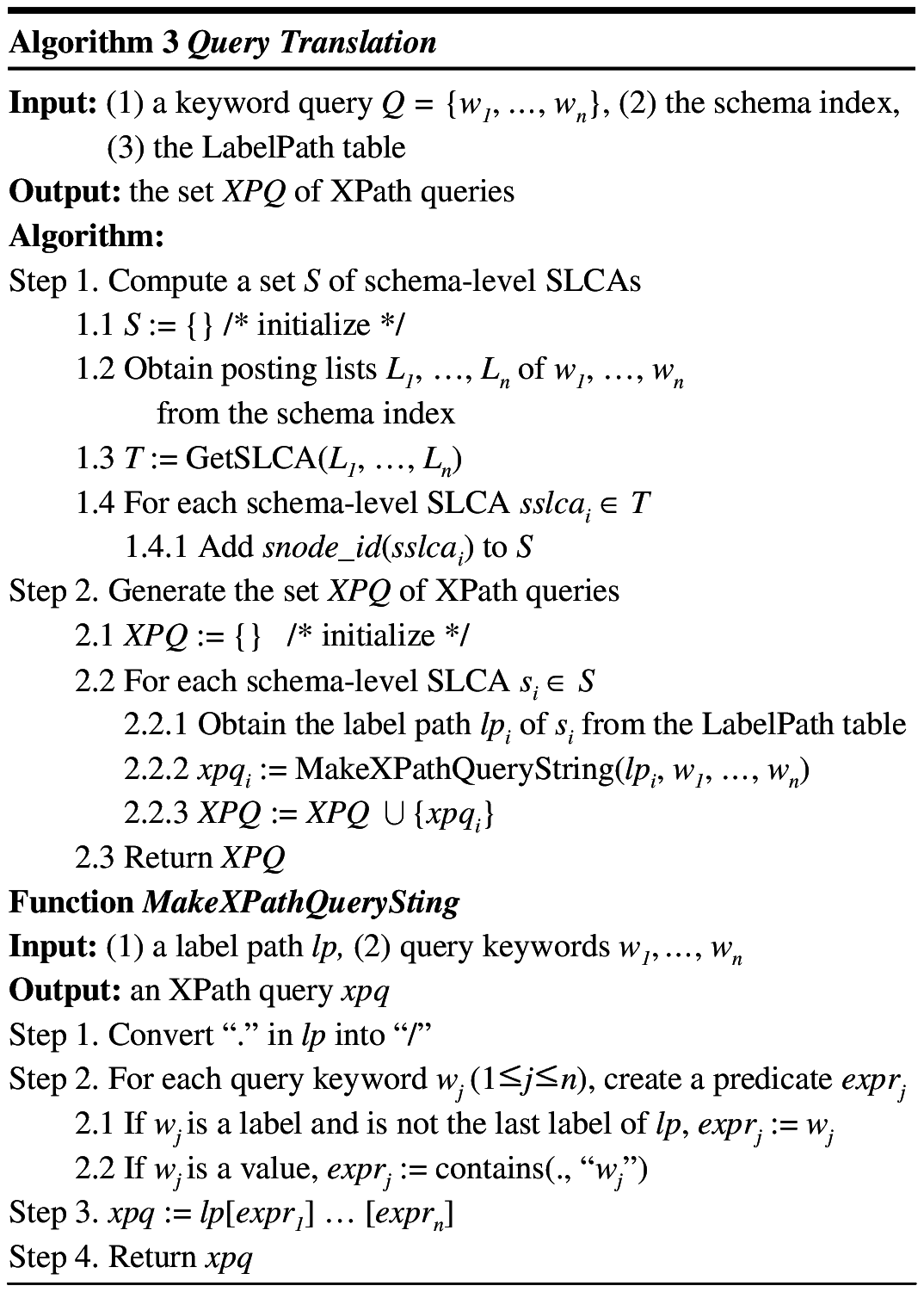}}
\ifx\vldbjformat\undefined \vspace*{-0.30cm} \fi
\caption{The query translation algorithm.}
\label{fig:QueryTranslation_Algorithm}
\end{figure}

We first compute schema-level SLCAs\,(or their ancestors) and then
generate XPath queries specifying their schema structures.
Fig.~\ref{fig:QueryTranslation_Algorithm} shows the algorithm {\it
Query Translation}, which consists of the following two steps.

In Step 1, we compute the set $S$ of schema-level SLCAs using the
GetSLCA function that implements the SLCA searching algorithm of
Xu and Papakonstantinou\,\cite{Xu05}. They use this function to compute instance-level SLCAs, but we use it here to compute schema-level ones. For
each schema-level SLCA $sslca_i$, we add the $snode\_id$ of $sslca_i$ to
$S$. In iterative $k$th-ancestor generalization, the algorithm is
modified to find ancestors of the schema-level SLCAs.

In Step 2, we generate an XPath query $xpq_i$ for each schema-level SLCA with the $snode\_id$ $s_i \in S$. In the XPath query generated from $s_i$, $s_i$ becomes the query result node and, at the same time, the branching query node since $s_i$ is a schema-level SLCA of all the query keywords; query keywords that are descendants of $s_i$ become the leaf query nodes. Here, we first obtain the label path $lp_i$ of $s_i$
by searching the LabelPath table using $snode\_id(s_i)$. We then make
the query string of $xpq_i$ by calling the MakeXPathQueryString
function with $lp_i$ and the query keywords. In Step 2.1 of the
MakeXPathQueryString function, we do not create a predicate when $w_i$ is
the last label of $lp$. It means that $w_i$ is the label of the schema-level SLCA. Since it is a part of $lp$ already, a predicate for it is
not needed.

\begin{figure}[h!]
\begin{center}
\begin{minipage}{3.5cm}
\begin{center}
\centerline{\includegraphics[width=3cm] {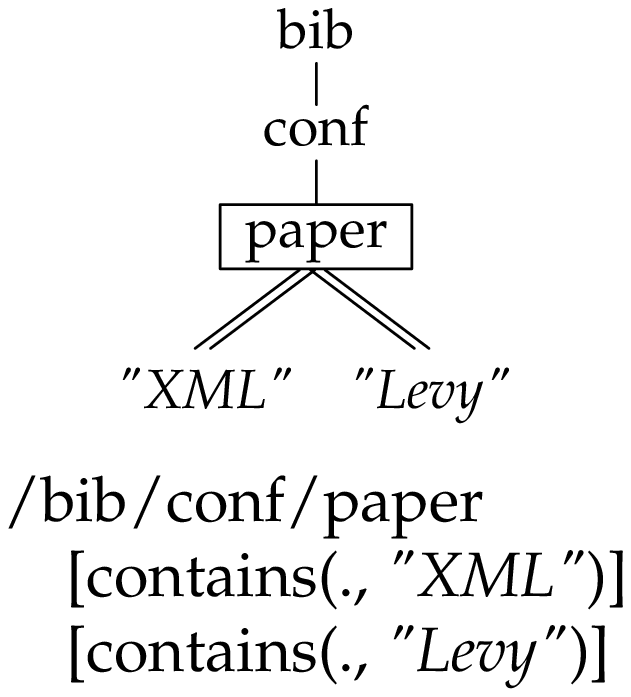}}
\ifx\vldbjformat\undefined \vspace{-0.4cm} \fi
{\footnotesize(a) $xpq_1$.}
\end{center}
\end{minipage}
\begin{minipage}{3.5cm}
\begin{center}
\centerline{\includegraphics[width=3cm] {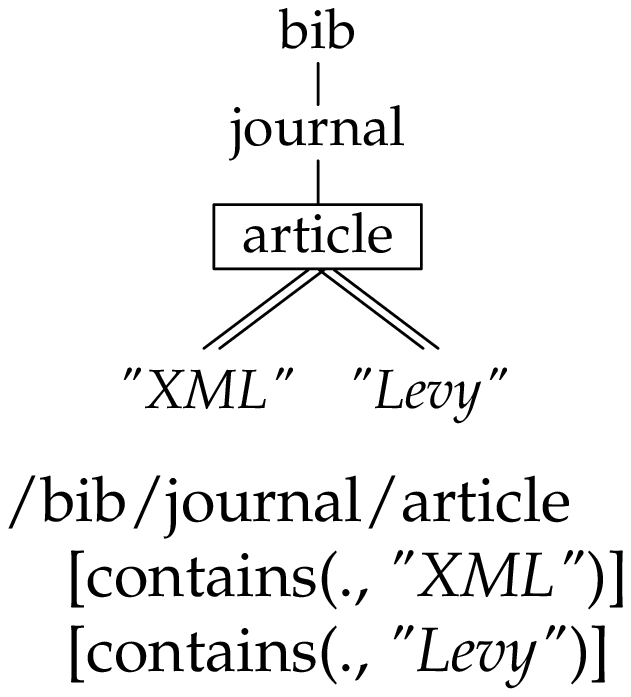}}
\ifx\vldbjformat\undefined \vspace{-0.4cm} \fi
{\footnotesize (b) $xpq_2$.}
\end{center}
\end{minipage}
\end{center}
\vspace{-0.2cm} \caption{The XPath queries generated from {\fssf``XML Levy"}.} \label{fig:XPath_queries}
\vspace{-0.2cm}
\end{figure}

\begin{example}
{\rm We translate a keyword query {\fssf ``XML Levy"} on the XML
data in Fig.~\ref{fig:motivating_example}(b) into XPath queries
$xpq_1$ and $xpq_2$ in Fig.~\ref{fig:XPath_queries} as follows. In Step
1, we first obtain the posting lists $L_1$, $L_2$ of {\fssf
``XML"}, {\fssf ``Levy"} by searching the schema index in
Fig.~\ref{fig:schema_index}. We then compute the set $T$ of
$numeric\_label \_path$'s of schema-level SLCAs for $L_1$ and $L_2$ by evaluating \ifx\vldbjformat\undefined GetSLCA($L_1$, $L_2$)\else GetSLCA ($L_1$, $L_2$)\fi. Here, $T$ = \{{\fssf ``0.1.6"},
{\fssf ``0.11.12"}\}. For each $sslca_i$ $\in$ $T$, we add
$snode\_id(sslca_i)$ to $S$. Thus, $S$ = \{{\fssf 6},
{\fssf 12}\} in Fig.~\ref{fig:dataguide}. In Step 2, for the schema-level SLCA with the $snode\_id$ $s_1 =$ {\fssf 6} $\in S$, we first obtain the label
path {\fssf ``bib.conf.paper"} of $s_1$ from the LabelPath table
 in Fig.~\ref{fig:labelpath_table}. We note that the
$label\_path\_id$ = $s_1$ = {\fssf 6}. We then create predicates
for {\fssf ``XML"} and {\fssf ``Levy"}. The predicates are ``{\fssf[contains(., ``XML")]}'' and ``{\fssf [contains(., ``Levy")]}''. Finally, we generate the XPath query $xpq_1$ by concatenating the label path and the predicates. We similarly generate
the XPath query $xpq_2$ for the schema-level SLCA with the $snode\_id$ $s_2$ = {\fssf 12}.}\hfill$\Box$
\end{example}

\subsubsection{Query Evaluation}\label{sec:query_evaluation}
The set of XPath queries obtained in the query translation step can be
evaluated with {\it any} existing XPath engine. In this section, we propose an efficient algorithm that simultaneously evaluates the specific set of XPath queries generated by our method.

In general, there are multiple structures matching the user's query intention, and thus, multiple XPath queries for those structures are generated from a keyword query. The result of the keyword query is the union of the results of these XPath queries. As explained in Section~\ref{sec:query_translation}, an XPath query $xpq_i$ generated from a schema-level SLCA $s_i$ has one branching node, i.e., $s_i$, and the label path of $s_i$ is the path from the root node to $s_i$. Query keywords that are descendants of $s_i$ become the leaf query nodes of $xpq_i$. The query $xpq_i$ finds the instance nodes that have the label path of $s_i$ and that contain all the query keywords (this is common to all $xpq_i$'s). We exploit this commonality for efficient simultaneous computation of multiple queries.

There has been a lot of work on XPath evaluation, but most of the work focuses on answering one query at a time. Some research efforts\,\cite{Brun03,Liu06a,Zhan08} have been done on answering multiple queries simultaneously, but they are not optimized for the specific set of XPath queries that are generated by our method.
Bruno et al.\,\cite{Brun03} and Zhang et al.\,\cite{Zhan08} only handle linear XPath queries. Liu et al.\,\cite{Liu06a} handle XPath queries with branches. This method is not suitable for the specific set of XPath queries because of the following reasons. They combine multiple queries into a single structure, called {\it super-twig query}, to exploit query commonalities. They only consider the scenario where query commonalities exist in the top parts---the parts close to the root node---of multiple original queries. However, in the specific set of XPath queries, much of the query commonalities exist in the bottom parts of the original queries, which consist of query keywords. Little query commonalities exist in the top parts since each query has a unique path from the root node to the branching node. Thus, in the worst case, the cost of the method is almost the same as that of processing one query at a time. In contrast, our algorithm simultaneously evaluates all the queries in this specific set by exploiting the query commonalities existing in the bottom parts of the original queries.




Since the queries in this specific set share the same query keywords that appear in the original keyword query, we can simultaneously evaluate all the queries by joining the posting lists of the query keywords. We obtain the posting lists from the instance index introduced in Section~\ref{sec:index_structure}. Suppose that XPath queries $xpq_1$,\,$xpq_2$,\,..., $xpq_m$ are obtained from a keyword query $Q$\,=\,\{$w_1$,\,$w_2$,\,..., $w_n$\}. We perform an index nested-loop join over the posting lists $L_j$ $(1$$\leq$$j$$\leq$$n)$ of query keywords $w_j$. For each posting in the outer-most posting list $L_1$, we identify the query to be evaluated from among $xpq_i$ $(1$$\leq$$i$$\leq$$m)$. Thus, we simultaneously evaluate different queries while we are scanning $L_1$. As explained in Section~\ref{sec:index_structure}, the posting of an instance node $o$ has the form $\langle$$inode\_id(o)$, $node\_path(o)$, $numeric\_label\_path(o)$$\rangle$ where $inode\_id(o)$ is the node id of $o$, $node\_path(o)$ the node path of $o$, and $numeric\_label\_path(o)$ the label path of $o$ that is represented as a sequence of integer ids rather than labels. $node\_path(o)$ contains the ids of the ancestor nodes of $o$ in the ascending order, and its last id is $inode\_id(o)$. A posting list is sorted in the ascending order of $inode\_id(o)$. Hereafter, we refer to an instance node $o$ by its posting for ease of exposition. For each posting $o_{1a}$ in $L_1$, we find the query to be evaluated using $numeric\_label\_path(o_{1a})$.
For $xpq_i$ $(1$$\leq$$i$$\leq$$m)$, if the path $p_i$ from the root node to the branching node of $xpq_i$ is a prefix of the label path of $o_{1a}$, $xpq_i$ must be the query that we need to evaluate for $o_{1a}$
since $xpq_i$ finds the instance nodes that have the label path $p_i$ and
that contain all the query keywords. Here, $o_{1a}$ matches the query keyword $w_1$ since $o_{1a}$ is a posting of $w_1$. We note that at most one $xpq_i$ is found since each query has a unique branching node. We compute the results only for the postings in $L_1$ that have the corresponding XPath query to be evaluated. Thus, we avoid unnecessary computation of spurious results. We note that, in contrast, the SLCA algorithm\,\cite{Xu05} computes SLCAs for all postings in $L_1$ incurring unnecessary computation.

We now explain how we evaluate $xpq_i$. Let $d_i$ be the depth of the branching node of $xpq_i$ from the root node, and $node\_path(o_{1a})[d_i]$ be the $d_i$th id of $node\_path(o_{1a})$. We need to check if the instance node $o$ with the id \ifx\vldbjformat\undefined $node\_path(o_{1a})[d_i]$ \else  $node\_path$ $(o_{1a})[d_i]$ \fi contains all the query keywords $w_j$ $(1$$\leq$$j$$\leq$$n)$. Here, $o$ corresponds to the query result since the branching node is the query result node in $xpq_i$. $o$ clearly contains $w_1$ since $o$ is an ancestor of $o_{1a}$. $o$ contains $w_j$ $(2$$\leq$$j$$\leq$$n)$ if there exists $o_{jb} \in L_j$ for each $L_j$ such that $node\_path(o_{jb})$ and $node\_path(o_{1a})$ have the same prefix from the root node to $d_i$. Since we assign a unique preorder id to each node in the XML data tree, $node\_path(o_{jb})$ and $node\_path(o_{1a})$ have the same prefix from the root node to $d_i$ if $node\_path(o_{jb})[d_i]$ = $node\_path(o_{1a})[d_i]$. Let $k$ be $node\_path(o_{1a})[d_i]$, which is $inode\_id(o)$. To check the existence of $o_{jb} \in L_j$ such that $node\_path(o_{jb})[d_i]$ = $k$, we utilize the subindex on $L_j$ whose key is $inode\_id$ of the posting in $L_j$, exploiting Lemmas~\ref{lemma:posting1} and \ref{lemma:posting2}. Here, we do not need to find all $o_{jb} \in L_j$ such that $node\_path(o_{jb})[d_i]$ = $k$ since we only need to check if $o$---which corresponds to the query result---contains $w_j$. By Lemmas~\ref{lemma:posting1} and \ref{lemma:posting2}, to
check the existence of $o_{jb} \in L_j$ such that $node\_path(o_{jb})[d_i]$ = $k$, we only need to find a posting $o_{jb}$ such that $inode\_id(o_{jb})$ is the smallest id that is greater than or equal to $k$ in $L_j$ and check whether $node\_path(o_{jb})[d_i]$ = $k$. In summary, we simultaneously evaluate all the queries $xpq_i$ $(1$$\leq$$i$$\leq$$m)$ through one scan of $L_1$ and an index nested-loop join over the posting lists $L_j$ $(1$$\leq$$j$$\leq$$n)$.

\begin{lemma}\label{lemma:posting1}
\ifx\vldbjformat\undefined \vspace{-0.4cm} \fi
{\rm  $inode\_id(o_{jb}) \geq k$ if
$node\_path(o_{jb})[d_i] = k$.}
\end{lemma}
\ifx\vldbjformat\undefined \vspace{-0.4cm} \else \vspace{-0.2cm} \fi
\noindent{\sc Proof:}
It is straightforward since we assign a preorder id to each node.
\hfill$\Box$

\begin{figure}[h]
\ifx\vldbjformat\undefined \vspace{0.2cm} \else \vspace{-0.3cm} \fi
\centerline{\includegraphics[width=2.5cm]
{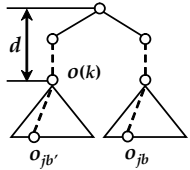}}
\ifx\vldbjformat\undefined \vspace{-0.6cm} \fi
\caption{An example XML data tree for the proof of Lemma~\ref{lemma:posting2}.}
\label{fig:o_jb}
\ifx\vldbjformat\undefined \vspace{-0.2cm} \fi
\end{figure}

\begin{lemma}\label{lemma:posting2}
\ifx\vldbjformat\undefined \vspace{-0.4cm} \fi
{\rm Let $inode\_id(o_{jb})$ be the smallest id that is greater than or equal to $k$ in $L_j$. If $node\_path(o_{jb})[d_i] \neq k$, then there is no $o_{jb'} \in L_j$ such that $node\_path(o_{jb'})[d_i] = k$.}
\end{lemma}
\ifx\vldbjformat\undefined \vspace{-0.4cm} \else \vspace{-0.2cm} \fi
\noindent{\sc Proof:}
Suppose that there exists $o_{jb'} \in L_j$ such that \ifx\vldbjformat\undefined $node\_path(o_{jb'})[d_i] = k$\else$node\_$ $path(o_{jb'})[d_i] = k$\fi. Then, as we see in Fig.~\ref{fig:o_jb}, $o_{jb'}$ must be in the subtree rooted at $o(k)$, and $o_{jb}$ must be in the right subtree of $o(k)$. Thus,
$inode\_id(o_{jb}) > inode\_id(o_{jb'}) \geq k$. This contradicts the assumption that $inode\_id(o_{jb})$ is the smallest id that is greater than or equal to $k$ in $L_j$.
\hfill$\Box$

Our algorithm uses the idea of XIR\,\cite{Park05} that exploits the schema information---more precisely, the label path---for XPath query processing.
XIR decomposes a given XPath query into linear XPath queries. A {\it linear XPath query}, which is also known as a {\it linear path expression}\,\cite{Park05}, is an XPath query without branches. It then finds a set of result node paths by processing each linear XPath query, and performs prefix match join between the sets of result node paths. Here, the {\it prefix match join}\,\cite{Park05} identifies the prefix (a subpath from the root to the branching node) of a node path on one side and finds the matching node paths having the same prefix on the other side of the join. In contrast to XIR, our algorithm simultaneously evaluates multiple XPath queries using the instance index without computing the result node paths a priori for each linear XPath query. In this sense, our algorithm is completely different from XIR.

\begin{figure}[!h]
\centerline{\includegraphics[width=8.5cm]{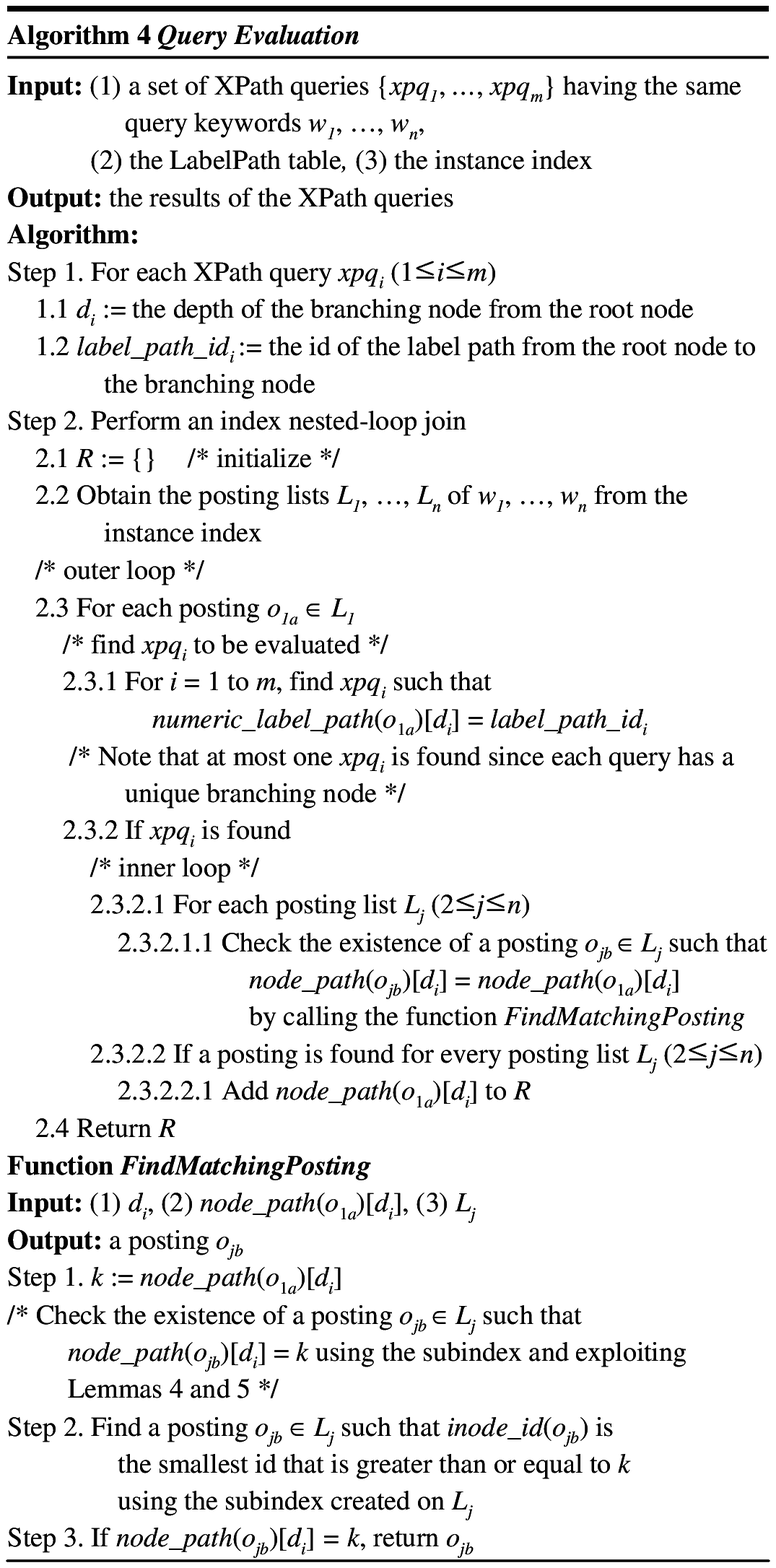}}
\ifx\vldbjformat\undefined \vspace*{-0.30cm} \fi
\caption{The query evaluation algorithm.}
\label{fig:QueryEvaluation_Algorithm}
\ifx\vldbjformat\undefined \vspace*{-0.20cm} \else \vspace*{-0.10cm} \fi
\end{figure}

Fig.~\ref{fig:QueryEvaluation_Algorithm} shows the query evaluation algorithm, which consists of the following two steps.

In Step 1, we obtain necessary information for query evaluation from the XPath queries. For each XPath query $xpq_i$ $(1$$\leq$$i$$\leq$$m)$, we first obtain the depth $d_i$ of the branching node from the root node (simply, the {\it branching depth}). We then obtain the id $label\_path\_id_i$ of the label path from the root node to the branching node using the LabelPath table.



In Step 2, we compute the results of the XPath queries. We first obtain the posting lists of the query keywords. We then scan the outer-most posting list $L_1$ and perform an index nested-loop join over the posting lists $L_j$ $(1$$\leq$$j$$\leq$$n)$. For each posting $o_{1a} \in L_1$, we find the query $xpq_i$ to be evaluated in Step 2.3.1.  If found, we do the inner loop step to check whether the node with the id $node\_path(o_{1a})[d_i]$ contains all the query keywords in Step 2.3.2.1. For each posting list $L_j$ $(2$$\leq$$j$$\leq$$n)$, we check the existence of $o_{jb} \in L_j$ such that $node\_path(o_{jb})[d_i]$ = $node\_path(o_{1a})[d_i]$, by calling the FindMatchingPosting function in Step 2.3.2.1.1. The FindMatchingPosting function finds such a posting using the subindex created on the posting list $L_j$ based on Lemmas~\ref{lemma:posting1} and \ref{lemma:posting2}. If a posting is found for every posting list $L_j$ $(2$$\leq$$j$$\leq$$n)$, we return $node\_path(o_{1a})[d_i]$ as the result of $xpq_i$.

Given a set of XPath queries \{$xpq_1$,\,$xpq_2$,\,..., $xpq_m$\} having the same query keywords \{$w_1$,\,$w_2$,\,..., $w_n$\}, the worst case time complexity $C_{XPath}$ of the query evaluation algorithm is $O(|L_1|(m + \sum^n_{j=2} log |L_j|))$ where $L_j$ $(1$$\leq$$j$$\leq$$n)$ is the posting list of $w_j$. For each posting in $L_1$, we find the query to be evaluated from among the $m$ queries and one posting from each of the other $n - 1$ posting lists. Finding a posting in $L_j$ using the subindex costs $O(log |L_j|)$.

We now compare the performance of our algorithm with that of the instance-level SLCA algorithm\,\cite{Xu05}. The worst case complexity of the SLCA algorithm is \ifx\vldbjformat\undefined $O(|L_1| d \sum^n_{j=2} log |L_j|)$\,\cite{Xu05} \else $O(|L_1| d \sum^n_{j=2} log$ $|L_j|)$\,\cite{Xu05} \fi where $d$ is the maximum depth of the XML data. In practice, $d$ of the SLCA algorithm and $m$ of our algorithm are small and do not affect performance significantly. Thus, the ``worst case'' performance of the two algorithms is almost the same. The critical benefit of our algorithm over the SLCA algorithm is that we avoid unnecessary computation of spurious results by only computing the results of the XPath queries obtained from schema-level SLCAs. This effect comes from the fact that we compute the results only for the postings in $L_1$ that have the corresponding XPath query to be evaluated (in Step 2.3.2) while the SLCA algorithm computes SLCAs for all postings in $L_1$.

\begin{figure}[h!]
\ifx\vldbjformat\undefined \else \vspace*{-0.40cm} \fi
\centerline{\includegraphics[width=8.5cm]{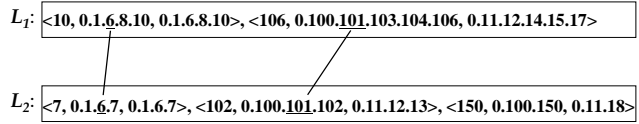}}
\ifx\vldbjformat\undefined\else \vspace*{0.20cm} \fi
\caption{An example of Algorithm 4.}
\label{fig:XPathProcessing_Example}
\ifx\vldbjformat\undefined \vspace*{-0.6cm} \else \vspace*{-0.2cm} \fi
\end{figure}

\begin{example}
{\rm We evaluate the XPath queries $xpq_1$ and $xpq_2$ in Fig.~\ref{fig:XPath_queries} as follows. In Step 1, the branching depth $d_i$ = {\fssf 3} for $xpq_i$ $(i = 1, 2)$. Since, in the LabelPath table in Fig.~\ref{fig:labelpath_table}, the id of the label path {\fssf ``bib.conf.paper"} is {\fssf 6} and that of \ifx\vldbjformat\undefined {\fssf ``bib.journal.article"} \else {\fssf ``bib.journal.} {\fssf article"} \fi is {\fssf 12}, $label\_path\_id_1$ = {\fssf 6} and $label\_path\_id_2$ = {\fssf 12}. In Step 2, we first obtain the posting lists $L_1$, $L_2$ of the query keywords {\fssf ``Levy"}, {\fssf ``XML"} as shown in Fig.~\ref{fig:XPathProcessing_Example}. For the posting $\langle inode\_id(o_{1a}), node\_path(o_{1a}), numeric\_label\_path(o_{1a})\rangle$ = $\langle${\fssf 10}, {\fssf 0.1.6.8.10}, {\fssf 0.1.6.8.10}$\rangle \in L_1$, $numeric\_label\_path(o_{1a})[d_1] = label\_path\_id_1$, or equivalently, {\fssf ``0.1.6.8.10"[3] = 6}. That is, {\fssf ``bib.conf.paper"} of $xpq_1$ is a prefix of the label path \ifx\vldbjformat\undefined {\fssf ``bib.conf.paper.author.ln"} \else {\fssf ``bib.conf.} {\fssf paper.author.ln"} \fi that corresponds to $numeric\_label\_path(o_{1a})$. Thus, $xpq_1$ is the query to be evaluated, and we do the inner loop step. We find a posting in $L_2$ such that \ifx\vldbjformat\undefined $node\_path(o_{2b})[d_1] = node\_path(o_{1a})[d_1]$ \else $node\_path(o_{2b})[d_1]$ $= node\_path(o_{1a})[d_1]$\fi = {\fssf ``0.1.6.8.10"[3]} = {\fssf 6} using the subindex created on $L_2$. Since there is a posting $\langle${\fssf 7}, {\fssf 0.1.6.7}, {\fssf 0.1.6.7}$\rangle \in L_2$ such that {\fssf ``0.1.6.7"[3] = 6}, we return {\fssf 6}, which is the node id of {\fssf paper(6)} in Fig.~\ref{fig:motivating_example}(b), as the result of $xpq_1$. For the posting $\langle${\fssf 106}, {\fssf 0.100.101.103.104.106}, {\fssf 0.11.12.14.15.17}$\rangle \in L_1$, we can similarly find the result {\fssf article(101)} of $xpq_2$.
}\hfill $\Box$
\end{example}

\section{Related Work}\label{sec:related_work}
There has been a lot of work on keyword search in relational databases\,\cite{Agra02,Bhal02,Hris02,Hris03a,Liu06,Luo07}, which inspired XML keyword search. However, the work on relational databases is not directly applicable to XML since the schema of XML data cannot always be mapped to a rigid relational schema\,\cite{Guo03} due to the semi-structured and heterogeneous nature of XML. Our approach
provides novel notions and algorithms that are suitable for the semi-structured and heterogeneous nature of XML and eliminates spurious results by exploiting the hierarchical nature of XML.

Extensive research has been done on XML keyword search. Under the
assumption that smaller subtrees are more relevant to the query,
most of the existing methods find the smallest subtrees containing
all the query keywords based on the concepts of the LCA or its
variants. Schmidt et al.\,\cite{Schm01} have introduced the notion of the
LCA, and Guo et al.\,\cite{Guo03} have defined a subset of LCAs and proposed an efficient ranking method for the subtrees rooted at the nodes in this set.
Xu and Papakonstantinou\,\cite{Xu08} have studied the properties of LCAs to accelerate the computation. Hristidis et al.\,\cite{Hris06} have focused on computing the whole subtrees rooted at LCAs. Xu and Papakonstantinou\,\cite{Xu05} have proposed the concept of the SLCA and presented algorithms for finding SLCAs efficiently. Sun et al.\,\cite{Sun07} have proposed a method that processes keyword queries involving boolean operators AND and OR under the SLCA semantics. Li et al.\,\cite{Li08} have proposed the concept of {\it Meaningful
LCA}\,({\it MLCA}), a concept similar to that of SLCA, and incorporated MLCA search in XQuery. Cohen et al.\,\cite{Cohe03} have attempted to find meaningful results based on a heuristic called {\it interconnection} relationship, and Li et al.\,\cite{Li07} have presented an efficient algorithm for the heuristic.

Liu and Chen\,\cite{Liu07} have pioneered a novel method for inferring {\it
return nodes} for XML keyword search. They have
proposed a system called {\it XSeek}, which infers desirable return
nodes by recognizing entities in the XML data. Huang et al.\,\cite{Huan08} have addressed the important problem of generating effective snippets (i.e., summaries) for XML search results. Liu and Chen\,\cite{Liu08} have proposed properties to find relevant nodes that matches query keywords in the subtree rooted at each SLCA. These schemes on generating return nodes are orthogonal to and can be incorporated into our method as we see in Section~\ref{sec:experiments}.

Several research efforts\,\cite{Cohe03,Li08,Yu07} have been made
to enable users to exploit partial knowledge of the schema in user
queries. The query models used in those methods are commonly called {\it labeled keyword search}\,\cite{Yu07}, which allows the user to annotate query keywords with labels. For example, in labeled keyword search, {\fssf``XML Levy"} is expressed as {\fssf``title:XML author:Levy"}. Using
this partial schema information, labeled keyword search can retrieve
more meaningful results than simple keyword search that specifies
only keywords. The search quality of labeled keyword search relies
on the correctness of the labels in a given query\,\cite{Li08}.
However, a casual user is unlikely to have perfect knowledge of
those labels\,\cite{Li08}. Our method does not have this problem
since it uses the simple keyword search model.

Yu and Jagadish\,\cite{Yu07} have proposed novel schema-based
matching methods for {\it labeled keyword search} and {\it Meaningful
Summary Query}\,(schema-aware query). They contrast with our framework
that supports schema-free keyword search. They use the schema of XML
data to define the matching semantics. In contrast, our method uses
the schema to efficiently resolve structural anomaly instead.

Most recently, Bao et al.\,\cite{Bao09} have proposed a probabilistic framework for inferring user's intention and ranking the query results. They compute the confidence level of each candidate {\it node type}, which is defined as a label path, using the statistics of the underlying XML data and use it to infer the user's intention. The method of Bao et al. processes queries at the instance level and additionally uses the schema to improve search quality. In contrast, our method, being primarily at the schema level, improves not only search quality using the schema but also search performance by processing queries at the schema level.

Besides, there has been extensive work done by W3C to define a
full-text extension of XQuery\,\cite{W3C}, which has today many
implementations such as GalaTex\,\cite{Gala}. Amer-Yahia et
al.\,\cite{Amer06} have presented efficient evaluation algorithms for
full-text XQuery queries, and Pradhan\,\cite{Prad06} has
demonstrated several optimization techniques. In this paper, our focus
is to effectively and efficiently support ``schema-free'' XML keyword
search where users only need to specify keywords as opposed to the
full-text extension of XQuery where users must specify structure
information as well as keywords according to the XQuery grammar.

There has been a lot of work on ranking schemes\,\cite{Agra02,Bao09,Bhal02,Guo03,Hris02,Hris03a,Hris03b,Li08b,Liu06,Tran09} for keyword search over XML, RDF, or relational databases. The ranking schemes and the concept of structural consistency can complement each other to help users find relevant results. For example, enforcing structural consistency could be too restrictive for certain applications, i.e, some query results eliminated by structural consistency may be relevant to the query. In this case, we can exploit structural consistency as one of the ranking criteria that measures the {\it meaningfulness}\,\cite{Yu07} of the results rather than as a criterion for removing spurious results as has similarly been suggested by Yu and Jagadish\,\cite{Yu07}.

\ifx\vldbjformat\undefined \else \vspace{-0.2cm} \fi
\section{Experimental Evaluation}\label{sec:experiments}
\subsection{Experimental Setup}
The goal of the experiments is to verify the advantage of our method in terms of search quality and search performance. As for {\it search quality}, we compare our method with SLCA\,\cite{Xu05} and MLCA\,\cite{Li08} as they are the state-of-the-art methods; we exclude XSEarch\,\cite{Cohe03} from the comparison since Li et al.\,\cite{Li08} have shown that MLCA is generally superior to XSEarch. As for {\it search performance}, we compare our method with SLCA, excluding MLCA from the comparison, since Xu and Papakonstantinou\,\cite{Xu05} have shown that the SLCA searching algorithm generally shows superior performance over the MLCA searching algorithm. In addition, we compare the index creation time and index size of our method with those of the SLCA method to show that an extra schema index for efficient structural consistency checking incurs negligible overhead to overall system performance. We use precision and recall as the measure for search quality. Following the common practice\,\cite{Cohe03,Li07,Li08}, we define the {\it desired results of a keyword query} as those returned by structured queries (XPath queries) corresponding to the keyword query, which are formulated by the users who participated in the experiments. We use the wall clock time as the measure for search performance and index creation, and the number of pages allocated for the index size.

Independent of the query processing method, we need to specify which
output\,(i.e., return nodes) generation strategies\,\cite{Liu07} to use: {\it Subtree Return}, {\it Path Return}, {\it Subtree-Entity Return}, and {\it
Path-Entity Return}. Subtree Return outputs the whole subtree rooted
at each query result. Path Return outputs the paths from the root of
each query result to the query keywords. Subtree-Entity Return and
Path-Entity Return first find the lowest entity ancestor-or-self node of each
query result, and then, output the subtree rooted at the node and the
paths from the node to the query keywords, respectively. In the same way as was done by Liu and Chen\,\cite{Liu07}, if a node with label $l_1$ has a one-to-many relationship with nodes with label $l_2$, we consider the nodes with label $l_2$ as entities. According to Liu and Chen\,\cite{Liu07}, Path Return usually has higher precision but lower recall than Subtree Return since it returns only paths. The strategies with entities generally have higher precision and recall than the ones without entities.

We present experimental results using the output strategies with entities since these strategies show superior search quality over those without. We note that this superiority has also been verified in all the experiments we performed. Thus, we omit experimental results for the output strategies without entities. For complete experimental results including other output strategies, please refer to our technical report\,\cite{Lee08b}. Hereafter, ``SC'' denotes our method; ``S-E'' a method with Subtree-Entity Return; and ``P-E'' a method with Path-Entity Return. For example, SC-S-E denotes our method with Subtree-Entity Return.

We have performed experiments using three real data sets and one synthetic data set. The first one is the DBLP data set\,\cite{Mikl}. We use the same schema used in the experiments by Xu and Papakonstantinou\,\cite{Xu05}, that groups the DBLP data set first by journal/conference names, and then, by years. The second one is the SIGMOD Record data set\,\cite{Mikl}. The third one is the NASA data set\,\cite{Mikl}, which consists of astronomical data. It has a complex and recursive schema and allows a wider variety of queries than the DBLP and SIGMOD Record data sets. The fourth and synthetic one is the XMark benchmark data set available at the XMark web site\,\cite{XMark}. These data sets have been extensively used in the existing work on XML keyword search\,\cite{Cohe03,Guo03,Hris06,Li07,Li08,Liu07,Schm01,Sun07,Xu05,Yu07}. Table~\ref{tbl:exp_data} shows statistics of these data sets. We see that the size of the schema is significantly smaller than that of the XML data.


\ifx\vldbjformat\undefined
\vspace*{-0.6cm}
\renewcommand{\baselinestretch}{1.0}
\fi

\ifx\vldbjformat\undefined \begin{table}[h] \else \begin{table*}[hbt] \fi
\begin{center}
\caption{Data statistics.}
\ifx\vldbjformat\undefined \vspace*{0.3cm} \fi
\label{tbl:exp_data}
{\small
\begin{tabular}{|c|c|c|c|c|c|}
\hline data set
& size & \# of instance nodes  & \# of distinct & \# of schema nodes & average \\
& & (excl. value nodes) & keywords & (excl. keywords) & depth
\\\hline \hline SIGMOD Record & 0.5 MBytes & 15,263 & 5,652 & 12 & 5
\\\hline DBLP & 127 MBytes & 3,736,406 & 572,062 & 145 & 3
\\\hline NASA & 23 MBytes & 530,528 & 48,430 & 110 & 6
\\\hline XMark & 111 MBytes & 2,048,193 & 127,905 & 548 & 5
\\\hline
\end{tabular}
}
\end{center}
\ifx\vldbjformat\undefined \vspace*{-0.6cm} \end{table} \else \vspace*{-0.6cm} \end{table*} \fi

\ifx\vldbjformat\undefined 
\else \vspace{0.3cm} \fi

\noindent{\bf Experiment 1:} To compare search performance and analyze the relationship between search performance and precision/recall in a controlled setting, we have generated the queries in Table~\ref{tbl:keyword_queries} for the DBLP, NASA, and XMark data sets\footnote{For the XMark data set, the XMark benchmark queries are not used since the queries are expressed in XQuery and has complex semantics such as path expressions, join, aggregation, grouping, and ordering. Since keyword queries have inherently limited expressive power, it is not feasible to rewrite all the benchmark queries into keyword queries. For some queries that do not have complex semantics and can easily be converted into keyword queries, e.g., $QX_4$ and $QX_7$, we exploit them.}. To show the cases where our method has low precision or recall, which are seldom, we add the following queries: $QD_6$, $QD_7$, $QX_6$, $QX_7$, $QN_4 \sim QN_7$. We also include $QD_8$, $QX_8$, $QN_8$ to test the case where users specify very long queries containing 9 $\sim$ 13 keywords. We run each query in Table~\ref{tbl:keyword_queries} ten times and measure precision, recall, and the average wall clock time. Since how the underlying XML data are stored highly affects the query result construction time, which is not our focus, we only access the root node $r$ of each query result and report the number of the descendant nodes of $r$ for the Subtree-Entity Return when measuring the wall clock time of query performance.



\begin{table}[h]
\begin{center}
\caption{Query sets.}
\ifx\vldbjformat\undefined \vspace*{0.3cm} \fi
\label{tbl:keyword_queries}
\begin{tabular}{|c|c|}
\hline ID & Query
\\ \hline \hline \multicolumn{2} {|c|}{DBLP data set}
\\ \hline \hline $QD_1$ & {\fssf``flexibility''}
\\ \hline $QD_2$ & {\fssf``scheduling management''}
\\ \hline $QD_3$ & {\fssf``quality analysis data''}
\\ \hline $QD_4$ & {\fssf``rule programming object system''}
\\ \hline $QD_5$ & {\fssf``Levy J Jagadish H''}
\\ \hline $QD_6$ & {\fssf``flexibility message scheme''}
\\ \hline $QD_7$ & {\fssf``ICDE XML Jagadish''}
\\ \hline $QD_8$ & {\fssf``distributed data base systems performance analysis}
\\ &  {\fssf Michael Stonebraker John Woodfill''}
\\ \hline \hline \multicolumn{2} {|c|}{NASA data set}
\\ \hline $QN_1$ & {\fssf``astroObjects''}
\\ \hline $QN_2$ & {\fssf``Michael magnitude''}
\\ \hline $QN_3$ & {\fssf``photometry galactic cluster Astron''}
\\ \hline $QN_4$ & {\fssf``pleiades dataset''}
\\ \hline $QN_5$ & {\fssf``PAZh components''}
\\ \hline $QN_6$ & {\fssf``pleiades journal''}
\\ \hline $QN_7$ & {\fssf``textFile name''}
\\ \hline $QN_8$ & {\fssf``accurate positions of 502 stars Eichhorn Googe }
\\ & {\fssf  Murphy Lukac''}
\\ \hline \hline \multicolumn{2} {|c|}{XMark data set}
\\ \hline $QX_1$ & {\fssf``Zurich''}
\\ \hline $QX_2$ & {\fssf``Arizona Mehrdad edu''}
\\ \hline $QX_3$ & {\fssf``Takano sun com  mailto''}
\\ \hline $QX_4$ & {\fssf``homepage name''}
\\ \hline $QX_5$ & {\fssf``Helena 96''}
\\ \hline $QX_6$ & {\fssf``mehrdad takano net''}
\\ \hline $QX_7$ & {\fssf``person id person0 name''}
\\ \hline $QX_8$ & {\fssf``harpreet mahony nodak edu 99 lazaro st el svalbard}
\\ & \fssf{ and jan mayen island''}
\\ \hline
\end{tabular}
\end{center}
\ifx\vldbjformat\undefined \vspace{-0.6cm} \else \vspace{-0.4cm} \fi
\end{table}

\ifx\vldbjformat\undefined \else \vspace{0.3cm} \fi
\noindent{\bf Experiment 2:}
To compare search performance for a real set of user queries, we have obtained
two hundred queries\footnote{For the list of queries, please
refer to \url{http://dblab.kaist.ac.kr/~drlee/sc.html}.} for each of the real data sets (a total of six hundred queries)---the DBLP, SIGMOD Record, and NASA data sets---from ten graduate students majoring in databases (but not involved in this project) for this purpose. We measure the wall clock time for all the queries.

\ifx\vldbjformat\undefined \else \vspace{0.3cm} \fi
\noindent{\bf Experiment 3:} To show the superiority of the query evaluation algorithm presented in Section~\ref{sec:query_evaluation}, we compare search performance of our method that uses the algorithm and the one that uses XIR\,\cite{Park05}, which does not process multiple XPath queries simultaneously. We measure the wall clock time for the six hundred queries used in Experiment 2.

\ifx\vldbjformat\undefined \else \vspace{0.3cm} \fi
\noindent{\bf Experiment 4:}
To compare search quality for real sets of user queries, we measure precision and recall for the six hundred queries used in Experiment 2.

\ifx\vldbjformat\undefined \else \vspace{0.3cm} \fi
\noindent{\bf Experiment 5:}
To compare the index creation time\footnote{In the index creation time, the time for XML document parsing, keyword extraction, and data loading is excluded.} and index size, we measure the wall clock time and the number of pages allocated.

\ifx\vldbjformat\undefined \else \vspace{0.3cm} \fi
\noindent{\bf Experiment 6:}
To test the scalability of our method, we generate XMark data sets by varying the size from 1 GBytes to 4 GBytes and from 100 MBytes to 10 GBytes. We measure the wall clock time for queries $QX_2$, $QX_3$, $QX_4$, and $QX_8$.

\ifx\vldbjformat\undefined \else \vspace{0.3cm} \fi
All the experiments are conducted on SUN Ultra 60 workstation with UltraSPARC-II 450MHz CPU and 512 MBytes of main memory. We implement all the methods on ODYSS-EUS ORDBMS\,\cite{Whan05}, which supports the inverted index. The page size for data and indexes is set to be 4096 bytes. We use the
Indexed Lookup Eager algorithm\,\cite{Xu05} as the SLCA searching algorithm since it generally shows superior performance over other algorithms. Finally, all the methods are implemented using C++.

\ifx\vldbjformat\undefined \else \vspace{-0.5cm} \fi
\subsection{Experimental Results}
\label{sec:exp_results}
\begin{figure*}[hbt]
\vspace{-0.2cm}
\centerline{\includegraphics[width=17cm]
{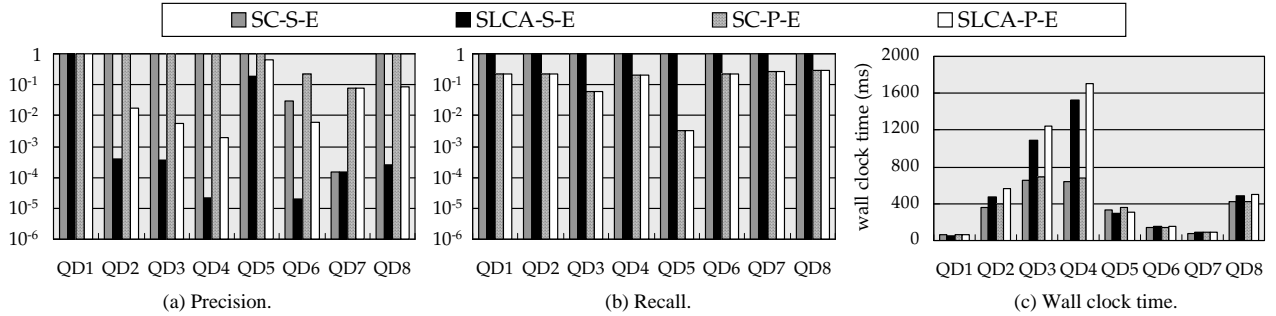}}
\caption{Precision, recall, and wall clock time of queries in Table~\ref{tbl:keyword_queries} for the DBLP data
set.}
\label{fig:dblp_result1}
\ifx\vldbjformat\undefined \vspace{-0.3cm} \else \vspace{-0.3cm} \fi
\end{figure*}

\ifx\vldbjformat\undefined \else \vspace{-0.3cm} \fi
\noindent{\bf Experiment 1:}
Fig.~\ref{fig:dblp_result1} shows the precision, recall, and wall clock time for the queries $QD_1 \sim QD_8$ in Table~\ref{tbl:keyword_queries} over the DBLP data set. SC-S-E (SC-P-E) improves the query performance by up to 2.4 times\,(2.5 times) over SLCA-S-E\,(SLCA-P-E). The reason for the improvement is that our method eliminates spurious results early by enforcing structural consistency at the schema-level. We note that the recall values of our method and SLCA are the same. The improvement becomes more marked when the precision of SLCA is low, i.e., when the number of spurious results is high. For example, in Fig.~\ref{fig:dblp_result1}(a), the precision of SLCA for $QD_4$ is lower than that for $QD_3$, and thus, in Fig.~\ref{fig:dblp_result1}(c), the query processing time for $QD_4$ is higher than that for $QD_3$, while those of our method are hardly changed. However, if the precision of SLCA is high, i.e., when there are few spurious results, for a specific query, our method could be marginally slower than SLCA due to the overhead of XPath query evaluation and iterative $k$th-ancestor generalization. For example, in Fig.~\ref{fig:dblp_result1}(c), our method is about 10\% slower than SLCA for $QD_1$ and $QD_5$.

In Fig.~\ref{fig:dblp_result1}(a), our method shows low precision for $QD_6$ and $QD_7$. For $QD_6$, there is a conference paper on {\fssf``flexibility message scheme"} in the database, but no journal article. In this case, our method finds spurious {\fssf journal} nodes through generalization, resulting in low precision. For $QD_7$, the user wants to find {\fssf``ICDE"} papers about {\fssf``XML"} authored by {\fssf``Jagadish"}, but our method and SLCA return the whole subtree rooted at {\fssf``ICDE"} conference node (or the paths from the conference node to the query keywords), resulting in the same low precision. Even for such queries, the precision of our method is higher than or equal to that of SLCA since our method is able to eliminate more spurious results than SLCA. For example, for $QD_6$, our method does not find spurious {\fssf conf} nodes since there is a paper on {\fssf``flexibility message scheme"}, but SLCA does.

The reason why the SLCA method often has very low precision is that it often finds more spurious SLCA nodes than correct ones. For example, there are only five publications of {\fssf``Levy"} on {\fssf``XML"} in the DBLP data set, but the SLCA method finds 50 SLCAs for the query {\fssf``XML Levy"}, 45 of which are spurious {\fssf conf} nodes. Furthermore, {\fssf conf} nodes typically include huge subtrees having thousands of nodes. Thus, the number of retrieved nodes that are spurious becomes very large leading to very low precision. The Subtree-Entity Return (S-E) has even lower precision because this strategy returns the whole subtree rooted at each query result, and the number of all nodes in the subtree is counted as the number of retrieved nodes.

\begin{figure*}[hbt]
\centerline{\includegraphics[width=17cm]
{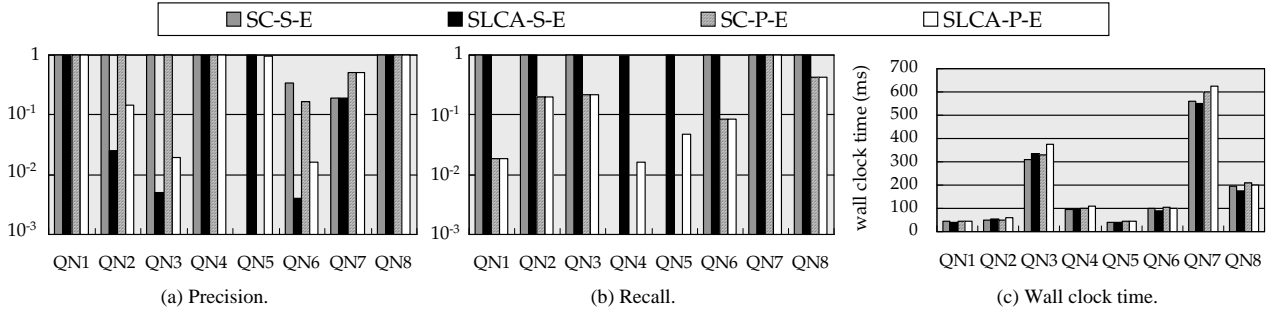}}
\ifx\vldbjformat\undefined \vspace{-0.2cm} \fi
\caption{Precision, recall, and wall clock time of queries in Table~\ref{tbl:keyword_queries} for the NASA data
set.}
\label{fig:nasa_result1}
\vspace{-0.1cm}
\end{figure*}

Fig.~\ref{fig:nasa_result1} shows the precision, recall, and wall clock time for the NASA data set, having a tendency similar to that of the DBLP data set except $QN_4$ and $QN_5$.

\begin{figure}[h]
\ifx\vldbjformat\undefined \else \vspace{-0.4cm} \fi
\centerline{\includegraphics[width=8cm]
{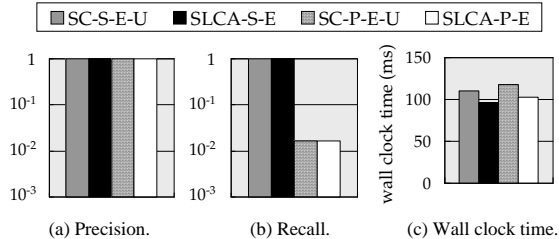}}
\ifx\vldbjformat\undefined \vspace{-0.2cm} \fi
\caption{Precision, recall, and wall clock time of $QN_4$ with relevance feedback.}
\label{fig:QN4}
\ifx\vldbjformat\undefined \else \vspace{-0.4cm} \fi
\end{figure}

For $QN_4$, the recall of our method, SC-S-E and SC-P-E, is almost 0 (both $1.3\times10^{-4}$ since they find the same {\fssf para} nodes). This is because the user intends to find more general results, which we regard as spurious results. For $QN_4$, {\fssf ``pleiades dataset"}, the user wants to find the subtrees rooted at {\fssf dataset} nodes that contain the keyword {\fssf ``pleiades"}. However, our method finds only the {\fssf para} nodes (i.e., paragraphs) that are contained in the subtrees rooted at the {\fssf dataset} nodes. Thus, we have very low recall. In contrast, the SLCA method finds (1) the {\fssf para} nodes and (2) the {\fssf dataset} nodes that do not have {\fssf para} nodes containing the keywords {\fssf ``pleiades"} and {\fssf ``dataset"}. (We note that the recall value of SLCA-S-E for $QN_4$ looks perfect in Fig.~\ref{fig:nasa_result1}(b), but it is not 1.0 since the SLCA method also finds the {\fssf para} nodes as our method does.) We can solve this low-recall problem using relevance feedback. The result is shown in Fig.~\ref{fig:QN4}. By using relevance feedback, we can generalize the {\fssf para} nodes to the {\fssf dataset} nodes and obtain the desired results.

\begin{figure}[h]
\ifx\vldbjformat\undefined \else \vspace{-0.4cm} \fi
\centerline{\includegraphics[width=8cm]
{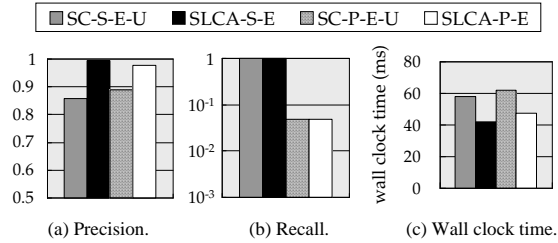}}
\ifx\vldbjformat\undefined \vspace{-0.2cm} \fi
\caption{Precision, recall, and wall clock time of $QN_5$ with relevance feedback.}
\label{fig:QN5}
\ifx\vldbjformat\undefined \else \vspace{-0.4cm} \fi
\end{figure}

For $QN_5$, the precision and recall of our method are both 0 constituting the worst case of our method. For $QN_5$, {\fssf ``PAZh components"}, the user wants to find the subtrees rooted at the {\fssf dataset} nodes that (1) have {\fssf altname} nodes whose value is {\fssf ``PAZh"} and (2) contain the keyword {\fssf ``components"}. However, our method finds {\fssf holding} nodes since there are {\fssf holding} nodes that contain the keywords {\fssf ``PAZh"} and {\fssf ``components"}. In contrast, existing methods find (1) the {\fssf holding} nodes and (2) the desired {\fssf dataset} nodes. We can also solve this problem by generalizing the {\fssf holding} nodes to the {\fssf dataset} nodes. The result is shown in Fig.~\ref{fig:QN5}. In Fig.~\ref{fig:QN5}(a), the precision of our method is worse than existing methods because we find spurious results during generalization as explained in Example~\ref{eg:low_recall} of Section~\ref{sec:comparision}\footnote{In Example~\ref{eg:low_recall}, {\fssf conf\_year} nodes correspond to {\fssf dataset} nodes; {\fssf chair} to {\fssf altname}; {\fssf ``Levy"} to {\fssf ``PAZh"}; {\fssf ``XML"} to {\fssf ``components"}; {\fssf paper} to {\fssf holding}.} while existing methods do not. That is, our method finds the {\fssf dataset} nodes that contain {\fssf ``PAZh"} and {\fssf ``components"} where the {\fssf altname} of the {\fssf dataset} node is not {\fssf ``PAZh"}.

\begin{figure*}[hbt]
\centerline{\includegraphics[width=17cm]
{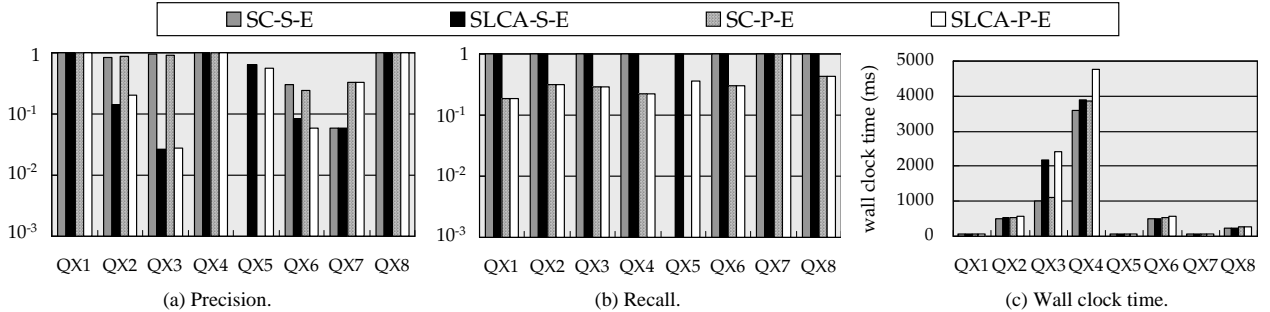}}
\vspace{-0.2cm}
\caption{Precision, recall, and wall clock time of queries in Table~\ref{tbl:keyword_queries} for the XMark data
set.}
\label{fig:xmark_result1}
\end{figure*}

\begin{figure}[h]
\ifx\vldbjformat\undefined \else \vspace{-0.4cm} \fi
\centerline{\includegraphics[width=8cm]
{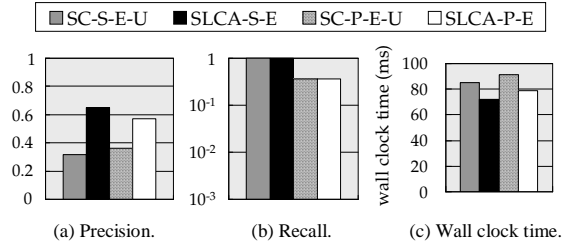}}
\ifx\vldbjformat\undefined \vspace{-0.2cm} \fi
\caption{Precision, recall, and wall clock time of $QX_5$ with relevance feedback.}
\label{fig:QX5}
\ifx\vldbjformat\undefined \else \vspace{-0.4cm} \fi
\end{figure}

Fig.~\ref{fig:xmark_result1} shows the precision, recall, and wall clock time for the XMark data set, showing a similar tendency to those of the DBLP and NASA data sets. Similar to $QN_5$ in the NASA dataset, $QX_5$ constitutes the worst case of our method. Fig.~\ref{fig:QX5} shows the results of $QX_5$ with relevance feedback.

\begin{figure*}[hbt]
\ifx\vldbjformat\undefined \else \vspace{-0.2cm} \fi
\begin{center}
\begin{minipage}{5.75cm}
\begin{center}
\centerline{\includegraphics[width=5.7cm]
{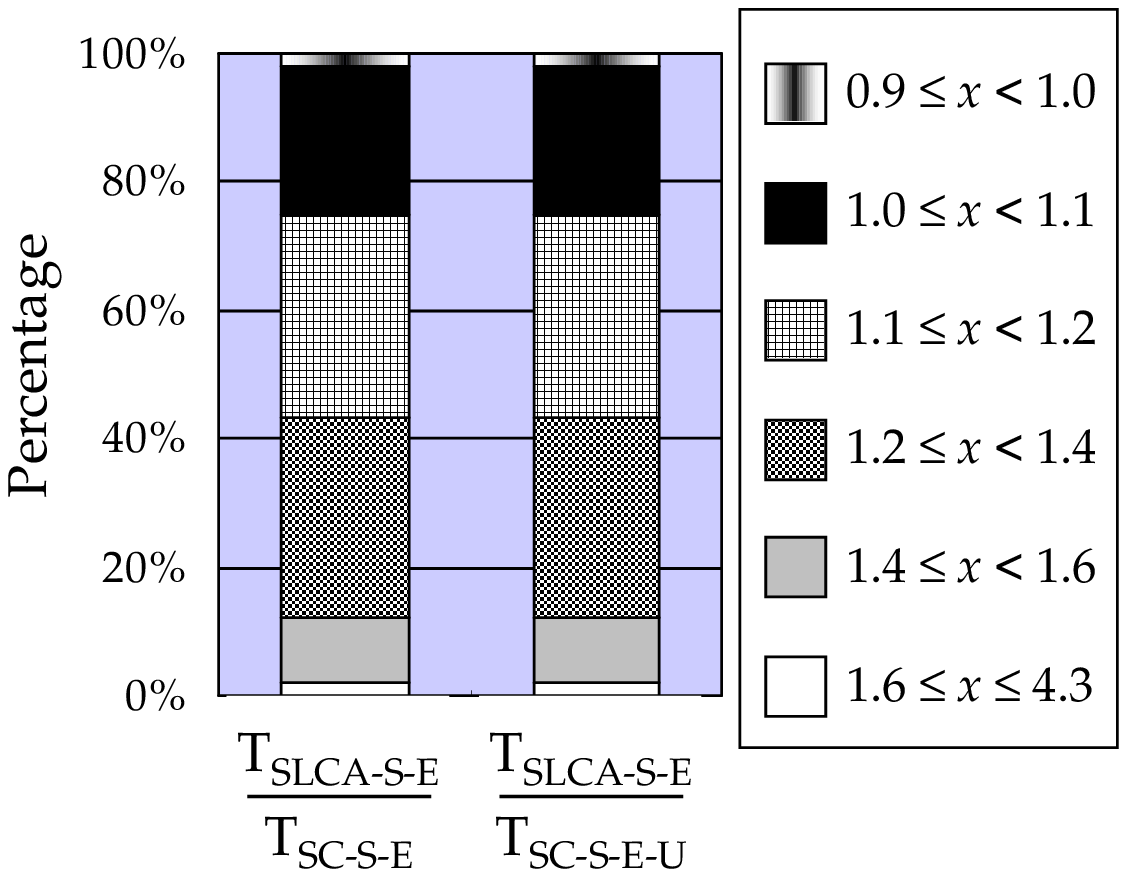}}
{\footnotesize (a) DBLP.}
\end{center}
\end{minipage}
\begin{minipage}{5.75cm}
\begin{center}
\centerline{\includegraphics[width=5.7cm]
{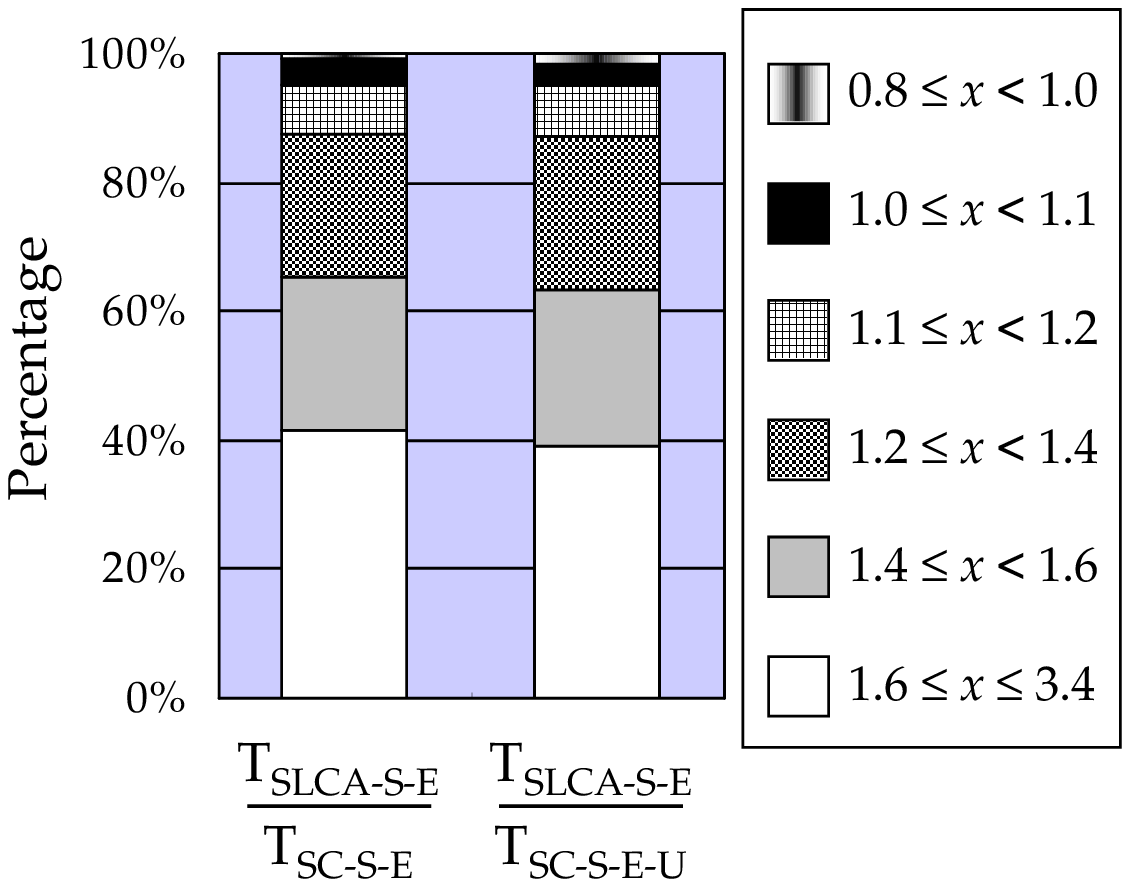}}
{\footnotesize (b) SIGMOD Record.}
\end{center}
\end{minipage}
\begin{minipage}{5.75cm}
\begin{center}
\ifx\vldbjformat\undefined \vspace{0.2cm} \fi
\centerline{\includegraphics[width=5.7cm]
{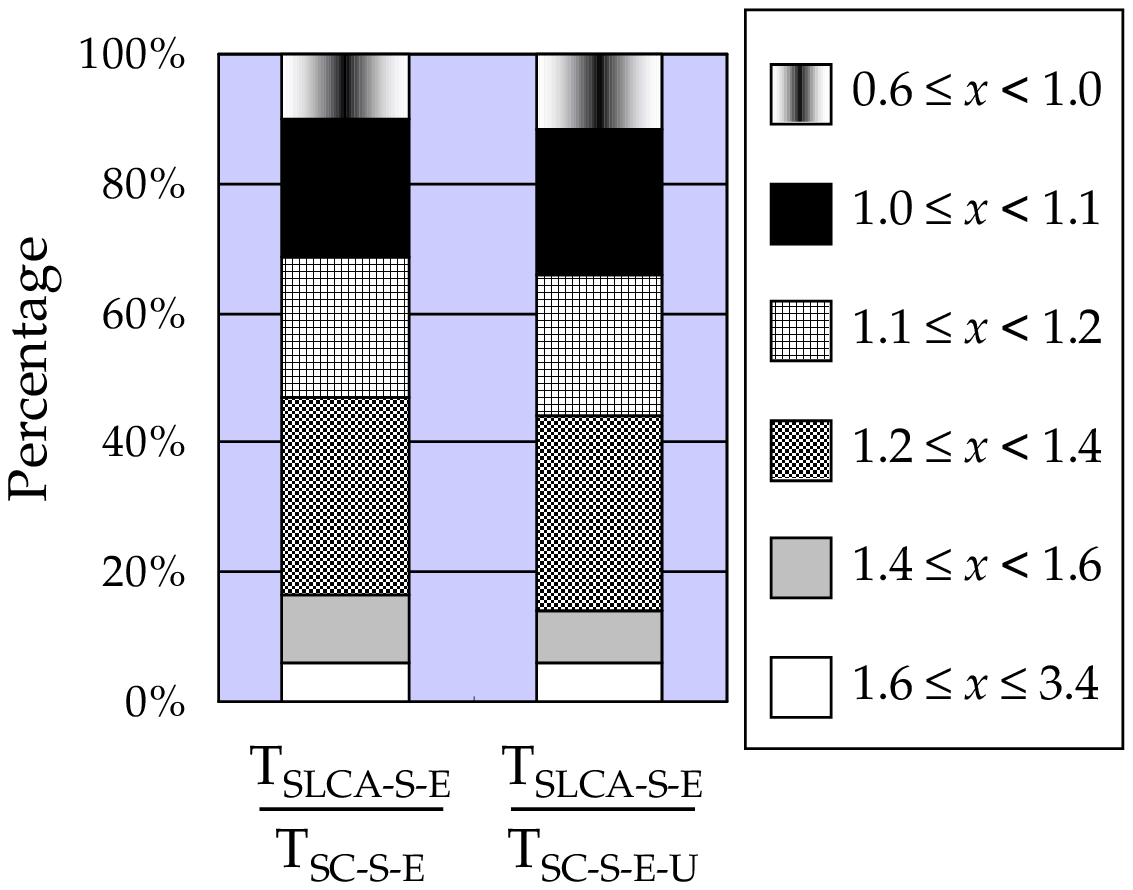}}
{\footnotesize (c) NASA.}
\end{center}
\end{minipage}
\end{center}
\vspace{-0.2cm}
\caption{The search performance results of six hundred queries for the DBLP, SIGMOD Record, and NASA data sets. The Y-axis represents the fraction of queries for which our algorithm has a given range of performance improvement over the SLCA algorithm.}
\label{fig:performance}
\vspace{-0.2cm}
\end{figure*}

\noindent{\bf Experiment 2:} Fig.~\ref{fig:performance} shows the search performance results for a real set of user queries. The Y-axis represents the fraction of queries for which our algorithm has a given range of performance improvement over the SLCA algorithm. The performance improvement is defined as the wall clock time $T_{SLCA-S-E}$ of SLCA over the wall clock time $T_{SC-S-E}$ of SC and denoted as $x$.
In Fig.~\ref{fig:performance}, ``-U'' denotes our method with relevance feedback. For the NASA data set in Fig.~\ref{fig:performance}(c), SC-S-E (SC-S-E-U) outperforms SLCA-S-E by more than 10\% for 69\% (66\%) of queries. In contrast, SLCA-S-E outperforms SC-S-E (SC-S-E-U) for only 10\% (12\%) of queries. Figs.~\ref{fig:performance}(a) and (b) show a tendency similar to that of the NASA data set. We omit the results for the Path-Entity Return (P-E) since they show a tendency similar to those of the Subtree-Entity Return (S-E).

\newpage
\noindent{\bf Experiment 3:} Our method that uses the algorithm presented in Section~\ref{sec:query_evaluation} outperforms the one that uses XIR\,\cite{Park05} by 1.8 $\sim$ 5.2 times since the algorithm simultaneously evaluates multiple XPath queries while XIR evaluates one query at a time.

\begin{figure*}[hbt]
\begin{center}
\begin{minipage}{8.5cm}
\begin{center}
\centerline{\includegraphics[width=8.2cm]
{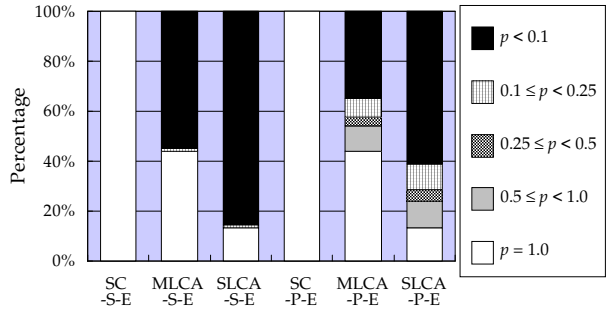}}
\ifx\vldbjformat\undefined \else \vspace{-0.1cm} \fi
{\footnotesize (a) Precision.}
\end{center}
\end{minipage}
\begin{minipage}{8.5cm}
\begin{center}
\centerline{\includegraphics[width=8.2cm]
{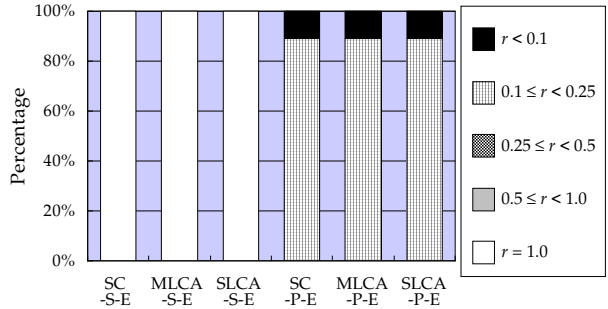}}
\ifx\vldbjformat\undefined \else \vspace{-0.1cm} \fi
{\footnotesize (b) Recall.}
\end{center}
\end{minipage}
\end{center}
\ifx\vldbjformat\undefined \vspace{-0.4cm} \else \vspace{-0.4cm} \fi
\caption{Precision and recall of two hundred queries for the DBLP data
set. The Y-axis represents the fraction of queries having a given precision/recall range.}
\label{fig:dblp_result2}
\end{figure*}

\begin{figure*}[hbt]
\begin{center}
\begin{minipage}{8.6cm}
\begin{center}
\centerline{\includegraphics[width=8.6cm]
{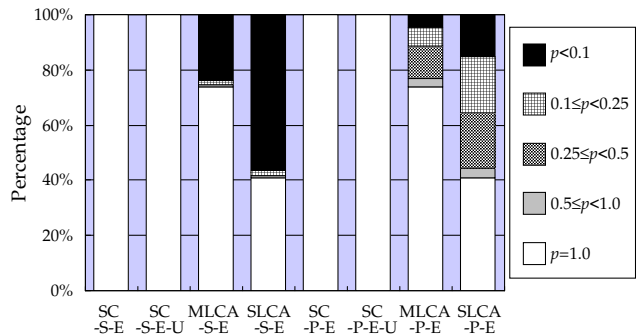}}
\ifx\vldbjformat\undefined \else \vspace{-0.1cm} \fi
{\footnotesize (a) Precision.}
\end{center}
\end{minipage}
\begin{minipage}{8.6cm}
\begin{center}
\centerline{\includegraphics[width=8.6cm]
{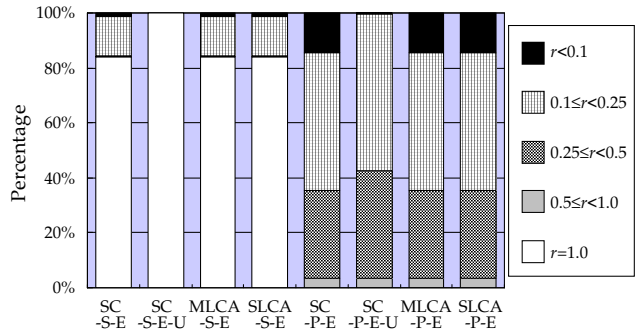}}
\ifx\vldbjformat\undefined \else \vspace{-0.1cm} \fi
{\footnotesize (b) Recall.}
\end{center}
\end{minipage}
\end{center}
\ifx\vldbjformat\undefined \vspace{-0.4cm} \else \vspace{-0.2cm} \fi
\caption{Precision and recall of two hundred queries for the SIGMOD Record data set. The Y-axis represents the fraction of queries having a given precision/recall range.}
\label{fig:sigmod_result2}
\ifx\vldbjformat\undefined \else \vspace{-0.2cm} \fi
\end{figure*}

\ifx\vldbjformat\undefined \else \vspace{0.2cm} \fi
\noindent{\bf Experiment 4:} Figs.~\ref{fig:dblp_result2} and \ref{fig:sigmod_result2} show the precision (denoted as $p$) and the recall (denoted as $r$) of two hundred queries over the DBLP data set and the SIGMOD Record data set, respectively. The Y-axis of the Figures represents the fractions of queries having given precision/recall ranges. MLCA and SLCA often find more spurious nodes than correct ones. For example, for the query {\fssf``activity recognition"}, they find 130 results, 122 of which are spurious {\fssf conf} or {\fssf journal} nodes. Thus, for the DBLP data set, the precision of SLCA and MLCA is less than 0.5 for 46\% $\sim$ 87\% of queries! For the SIGMOD Record data set, their precision is less than 0.5 for 23\% $\sim$ 59\% of queries. In contrast, the precision of our method is 1.0 for all queries since it eliminates all the spurious results by enforcing structural consistency. We note that the recall values of our method, MLCA, and SLCA are the same. These results are similar to those of Experiment 1.

In Fig.~\ref{fig:sigmod_result2}(b), SC-S-E, MLCA-S-E, and SLCA-S-E show low recall for about 16\% of queries. In this case, the users want the articles of an author, e.g., {\fssf ``Jennifer Widom"}, but all methods return the author in the articles since the author is the lowest entity containing all the query keywords. However, SC-S-E-U shows perfect recall since it finds the articles of an author by using relevance feedback. The average number of relevance feedbacks provided by the users for the 200 queries on the SIGMOD Record data set is 0.36/query.

\begin{figure*}[hbt]
\begin{center}
\begin{minipage}{8.6cm}
\begin{center}
\centerline{\includegraphics[width=8.6cm]
{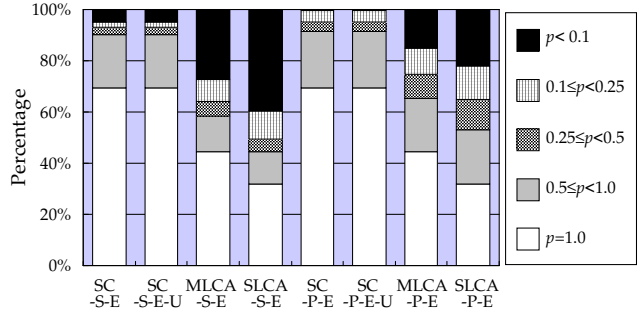}}
\vspace{-0.1cm}
{\footnotesize (a) Precision.}
\end{center}
\end{minipage}
\begin{minipage}{8.6cm}
\begin{center}
\centerline{\includegraphics[width=8.6cm]
{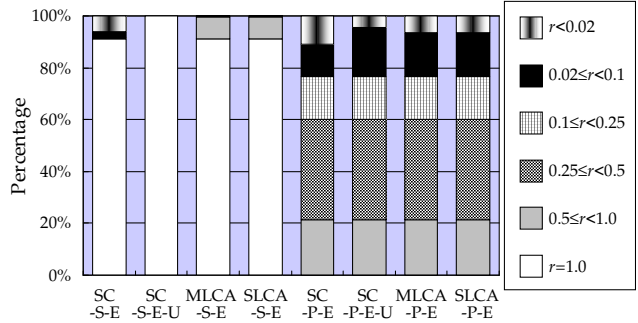}}
\vspace{-0.1cm}
{\footnotesize (b) Recall.}
\end{center}
\end{minipage}
\end{center}
\ifx\vldbjformat\undefined \vspace{-0.4cm} \else \vspace{-0.2cm} \fi
\caption{Precision and recall of two hundred queries for the NASA data
set. The Y-axis represents the fraction of queries having a given precision/recall range.}
\label{fig:nasa_result2}
\ifx\vldbjformat\undefined \else \vspace{-0.1cm} \fi
\end{figure*}

Fig.~\ref{fig:nasa_result2} shows the precision and the recall of two hundred queries over the NASA data set. The precision of SLCA and MLCA is less than 0.5 for 35\% $\sim$ 56\% of queries. In contrast, the precision of our method is less than 0.5 for only 9\% $\sim$ 10\% of queries. Here, our method shows low precision for some queries due to the complex schema of the NASA data set. For example, for the query {\fssf ``radio journal"}, the user wants to find journal articles on {\fssf ``radio"}. Our method finds not only correct results but also spurious results such as {\fssf revision} nodes, as SLCA and MLCA do, since there are {\fssf revision} nodes that contain the keywords {\fssf ``radio"} and {\fssf ``journal"}.

In Fig.~\ref{fig:nasa_result2}(b), for about 9\% of queries, the recall values of our method without relevance feedback are lower than those of SLCA and MLCA due to the same reason as in Example~\ref{eg:low_recall} of Section~\ref{sec:comparision}. However, by using the relevance feedback, we can archive higher recall values than SLCA and MLCA. The average number of relevance feedbacks provided by the users for the 200 queries on the NASA data set is 0.30/query.

\begin{figure}[h]
\centerline{\includegraphics[width=8cm]
{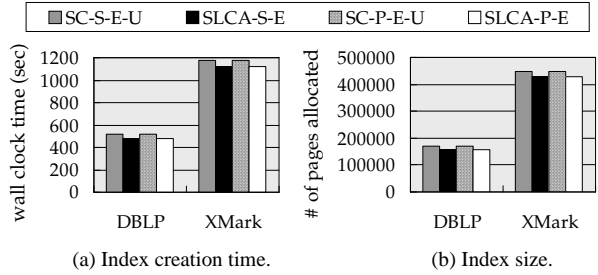}}
\caption{Index creation time and index size for the DBLP and XMark
data sets.}
\label{fig:dblp_xmark_index}
\ifx\vldbjformat\undefined \vspace{-0.6cm} \else \vspace{-0.4cm} \fi
\end{figure}

\ifx\vldbjformat\undefined \else \vspace{0.2cm} \fi
\noindent{\bf Experiment 5:}
Fig.~\ref{fig:dblp_xmark_index} shows the index creation time and
the index size. All methods use an inverted index for XML data and the {\it Dewey index}\,\cite{Liu07} to find the lowest entity ancestor of each query result. SC-S-E and SC-P-E additionally use the schema index for efficient structural consistency checking. Thus, the index creation time of SC-S-E and SC-P-E is about 5\% $\sim$  7\% longer, and the index size is about 5\% $\sim$ 7\% larger than those of SLCA-S-E and SLCA-P-E. This verifies that an extra schema index incurs negligible overhead to overall system performance.
We note that the index is bigger than the original data due to the space required for storing id paths from the root to each node. SLCA-based methods have similar space overhead since they also use id paths, i.e., Dewey numbers. We could reduce the space by exploiting the UTF-8 encoding as an efficient way to represent id paths, which was proposed by Tatarinov et al.\,\cite{Tata02}.

\begin{figure*}[hbt]
\ifx\vldbjformat\undefined \vspace{0.3cm} \fi
\centerline{\includegraphics[width=17.8cm]
{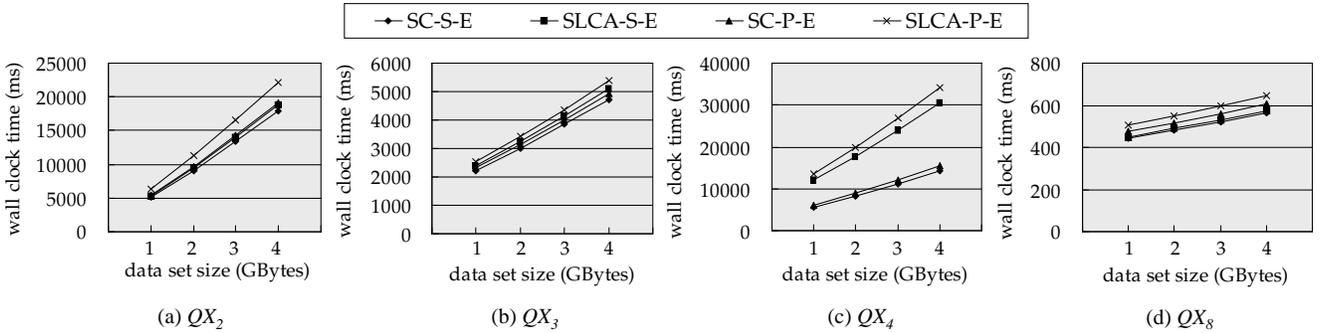}}
\caption{Query processing time with increasing data set size from 1 GBytes to 4 GBytes in a linear scale.}
\label{fig:scalability}
\end{figure*}

\begin{figure*}[hbt]
\centerline{\includegraphics[width=17.8cm]
{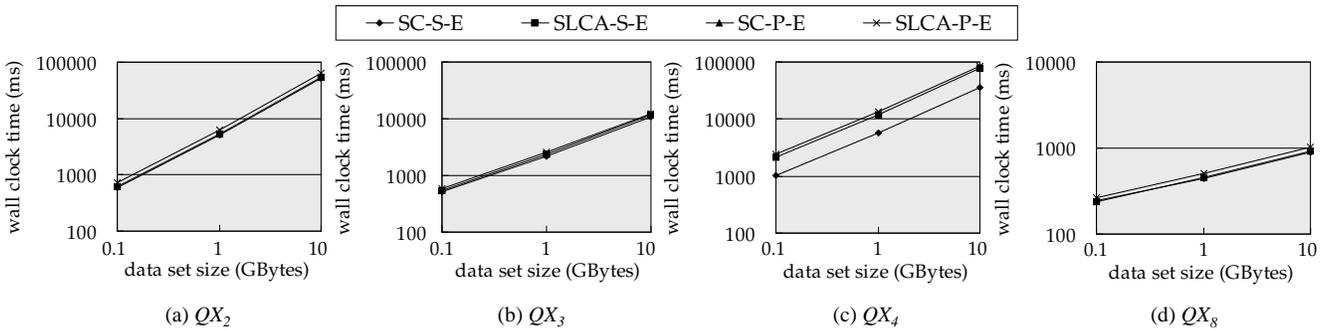}}
\caption{Query processing time with increasing data set size from 100 MBytes to 10 GBytes in a logarithmic scale.}
\label{fig:scalability2}
\end{figure*}

\ifx\vldbjformat\undefined \vspace{0.4cm} \else \vspace{0.2cm} \fi
\noindent{\bf Experiment 6:}
Figs.~\ref{fig:scalability} and \ref{fig:scalability2} show the processing time of queries $QX_2$, $QX_3$, $QX_4$, and $QX_8$ as the data set size is varied from 1 GBytes to 4 GBytes and from 100 MBytes to 10 GBytes. As we can see, the processing time of all methods increases approximately linearly when the data set size increases and that our methods are largely superior or comparable to SLCA-based methods.

\section{Conclusions}\label{sec:conclusions}
\ifx\vldbjformat\undefined \vspace{-0.4cm} \fi
We have proposed a new notion of structural consistency (and structural anomaly) in XML keyword search. By exploiting structural
consistency, we can eliminate spurious results having the same result structure consistently. We have introduced the concept of the result structure in
Definition~\ref{def:result_structure} and the smallest result
structure in Definition~\ref{def:smallest_result_structure}. We have formally defined the structural anomaly in Definition~\ref{def:structural_consistency} as a phenomenon where there exist result structures that structurally contain other result structures. We have defined the structural consistency as a property where there is no structural anomaly in the query results.

We have proposed a naive algorithm that resolves structural anomaly at
the {\it instance} level. We have then proposed an advanced algorithm
that resolves structural anomaly at the {\it schema} level. To this
end, we have formally analyzed the relationship between the set of schema-level SLCAs and the set of instance-level SLCAs in Lemmas~\ref{lemma:preceq} $\sim$ \ref{lemma:kth_ancestor}, identified the discrepancies between them, and proposed the notion of iterative $k$th-ancestor generalization to resolve the anomalies (false dismissal and phantom schema structures) that
are caused by these discrepancies. We have formally proved that the
proposed algorithms produce the same set of results preserving structural consistency in Theorem~\ref{theorem:schema_level}.
We have proposed a solution using relevance feedback for the problem where our method has low recall; this problem occurs when it is not the user's intention to find more specific results. We have provided an efficient algorithm that simultaneously evaluates multiple XPath queries generated by our method. We have implemented our method in a full-fledged object-relational DBMS.

We have performed extensive experiments using real and synthetic data
sets. Experimental results show that our method improves precision significantly compared with the existing methods while providing comparable recall for most queries. Experimental results also show that our method improves the query performance over the existing methods significantly by removing spurious results early.

\ifx\vldbjformat\undefined \else \vspace{-0.4cm} \fi
\section*{Acknowledgements}
\vspace{-0.2cm}
Earlier versions of this paper were presented in the KAIST CS technical reports\,\cite{Lee08b,Lee07} and in the PhD dissertation\,\cite{Lee08a} of Ki-Hoon Lee. This research was partially supported by the National Research Lab Program through the National Research Foundation of Korea (NRF) funded by the Ministry of Education, Science and Technology (No. R0A-2007-000-20101-0).
This work was also partially supported by the Internet Services Theme Program funded by Microsoft Research Asia and by the KAIST-Microsoft Research Collaboration Center (KMCC).

\ifx\vldbjformat\undefined \else \vspace{0.1cm} \fi
\noindent{\large\bf Appendix~A. Proof of Lemma~\ref{lemma:preceq}}\\
{\small
Let \{$w_1$, $w_2$, ..., $w_n$\} be the set of query keywords of $Q$, and $l_1$.$l_2$.$\cdots$.$l_m$ be the incoming label path of $srs_i$. We need to show that there always exists a schema-level SLCA $s$ such that $l_1$.$l_2$.$\cdots$.$l_m$ is a prefix of the label path of
$s$. Since $srs_i$ is a smallest result structure of instance-level
SLCAs, there exists an instance node $v$ such that $l_1$.$l_2$.$\cdots$.$l_m$ is the label path of $v$, and $w_1$, $w_2$, ..., $w_n$ are descendants of $v$. It follows that there exists a schema node $s_a$ such that $l_1$.$l_2$.$\cdots$.$l_m$ is the label path of $s_a$ and $w_1$, $w_2$, ..., $w_n$ are descendants of $s_a$ (i.e., $srs_i \equiv ss(s_a)$) since the DataGuide$^+$ has every unique label path of instance nodes. Thus, by the definition of schema-level SLCA, there exists a schema-level SLCA $s$ such that $ss(s_a)$\,$\preceq$\,$ss(s)$.\hfill$\Box$
}

\ifx\vldbjformat\undefined \else \vspace{0.1cm} \fi
\noindent{\large\bf Appendix~B. Proof
of Lemma~\ref{lemma:kth_ancestor}}\\
{\small
Let $ILP(srs_i)$ be the incoming label path of $srs_i$, and $ILP(ss_j)$ be the incoming label path of $ss_j$. Since $srs_i$\,$\prec$\,$ss_j$, $ILP(srs_i)$ is a proper prefix of $ILP(ss_j)$. This implies that there must exist a $k$th-ancestor $s_a$ $(1$\,$\leq$\,$k$\,$\leq$\,$depth(s))$ of the schema-level SLCA $s$ whose label path is the same as $ILP(srs_i)$. Here, $ss(s_a)$\,$\equiv$\,$srs_i$ since the label path of $s_a$ is the same as $ILP(srs_i)$.\hfill $\Box$
}

\end{document}